\documentclass[pdflatex,sn-mathphys-num]{sn-jnl}

\usepackage{graphicx}%
\usepackage{multirow}%
\usepackage{amsmath,amssymb,amsfonts}%
\usepackage{amsthm}%
\usepackage{mathrsfs}%
\usepackage[title]{appendix}%
\usepackage{xcolor}%
\usepackage{textcomp}%
\usepackage{manyfoot}%
\usepackage{booktabs}%
\usepackage{placeins}
\usepackage{algorithm}%
\usepackage{algorithmicx}%
\usepackage{algpseudocode}%
\usepackage{listings}%
\usepackage{array} 

\theoremstyle{thmstyleone}%
\theoremstyle{thmstyletwo}%

\theoremstyle{thmstylethree}%

\begin{document}

\title[Article Title]{QAOA-Predictor: 

Forecasting Success Probabilities and Minimal Depths for Efficient Fixed-Parameter Optimization}


\author*[1]{\fnm{Rodrigo} \sur{Coelho}} \email{rodrigo.coelho@iisb.fraunhofer.de}

\author[1,2]{\fnm{Georg} \sur{Kruse}} \email{georg.kruse@iisb.fraunhofer.de}

\author[3]{\fnm{Jeanette Miriam}\sur{Lorenz}} \email{jeanette.miriam.lorenz@iks.fraunhofer.de}

\affil[1]{\orgname{Fraunhofer IISB}, \orgaddress{\street{Schottkystraße 10}, \city{Erlangen}, \postcode{91054}, \state{Bavaria}, \country{Germany}}}

\affil[2]{\orgname{Technical University Munich}, \orgaddress{\street{Arcisstraße 21}, \city{Munich}, \postcode{80333}, \state{Bavaria}, \country{Germany}}}

\affil[3]{\orgname{Fraunhofer IKS}, \orgaddress{\street{Hansastraße 32}, \city{Munich}, \postcode{80686}, \state{Bavaria}, \country{Germany}}}


\abstract{Quantum Computing promises to solve complex combinatorial optimization problems more efficiently than classical methods, with the Quantum Approximate Optimization Algorithm (QAOA) being a leading candidate. Recent fixed-parameter variations of QAOA eliminate costly run-time optimization, but determining their optimal initialization as well as the number of required layers ($p$) for a target solution remains a critical, unsolved challenge. In this work, we propose a novel approach using a Graph Neural Network (GNN) to predict QAOA performance: Based on a graph representation of the problem, the GNN forecasts the probability of the optimal solution in the resulting distribution across different parameter initializations and layer depths for a wide variety of combinatorial optimization problems. We demonstrate that the GNN accurately predicts QAOA performance within a $10\%$ margin of the true values. Furthermore, the model exhibits strong generalization capabilities across unseen problem classes, larger problem sizes, and higher layer counts. Our approach allows to identify viable problem instances for QAOA and to select an adequate parameter initialization strategy with minimal layer depth, without the need of costly parameter optimization.}




\maketitle

\section{Introduction}\label{sec1}

\begin{figure}
    \centering
    \includegraphics[width=\linewidth]{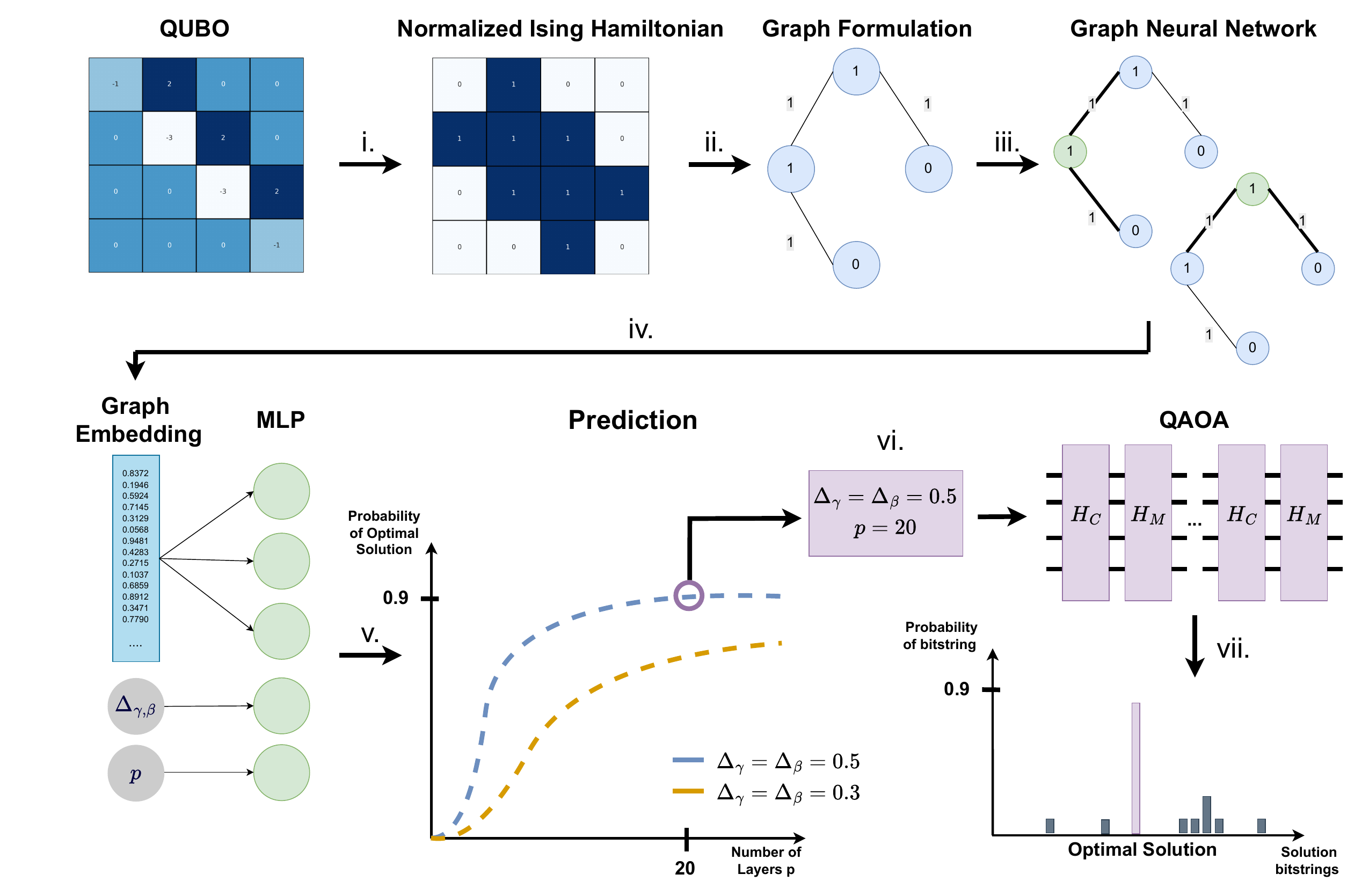}
    \caption{Overview of our method: For a specific problem, its QUBO matrix is generated, which is subsequently transformed into a (normalized) Ising Hamiltonian (i). The Hamiltonian matrix is then re-formulated as a graph (ii), which is fed to a Graph Neural Network (GNN) (iii) that outputs a graph-embedding (iv). Then, a Multi-Layer Perceptron (MLP) receives this graph-embedding together with values of $\Delta_{\gamma,\beta}$ and the numbers of layers $p$, and predicts the performance of LR-QAOA for this problem instance (v), measured as the probability of the optimal solution in the resulting probability distribution. This prediction can than be used to initialize LR-QAOA with the respective parameters (vi) and sample the optimal solution with the predicted probability (vii) without the need of additional parameter optimization.}
    \label{fig:method}
\end{figure}

Quantum computers hold the potential to outperform classical computers in solving a wide range of challenging problems. A prominent example is Shor's algorithm \cite{shor1999polynomial}, which achieves a super-polynomial speedup in factoring large integers compared to the best-known classical algorithms. This capability poses significant threats to current public-key cryptography systems \cite{mavroeidis2018impact}.

Another prominent application domain for quantum computing is combinatorial optimization. Many practically relevant problems in this class are NP-hard, implying that no known classical algorithm achieves polynomial scaling in the worst case. Consequently, the computational cost of the best available classical methods typically grows exponentially with problem size, rendering large-scale instances intractable on current and foreseeable classical hardware.

To overcome these limitations, quantum algorithms have been developed that exploit superposition and entanglement to explore solution spaces more efficiently. Notable approaches include adiabatic quantum computation (e.g. Quantum Annealing) \cite{yarkoni2022quantum} and variational quantum algorithms \cite{cerezo2021variational}. The later are hybrid quantum-classical methods, which partly can operate efficiently on near-term hardware. Among these, the Quantum Approximate Optimization Algorithm (QAOA) \cite{farhi2014quantum} stands out as a leading candidate. QAOA is specifically tailored for binary combinatorial optimization problems, particularly those expressed as Quadratic Unconstrained Binary Optimization (QUBO) instances \cite{glover2018tutorial}. The algorithm constructs a parameterized quantum circuit consisting of $p$ alternating layers: one applying the problem Hamiltonian (encoding the objective function) and the other a mixing Hamiltonian (typically transverse-field drivers to explore the solution space). By classically optimizing the parameters of these layers ($\gamma$ and $\beta$ for problem and mixing Hamiltonian respectively) to minimize the expectation value of the problem Hamiltonian, QAOA prepares states that yield high-quality approximate solutions upon measurement.

However, the iterative classical feedback loop used to optimize the parameters of the layers introduces two crucial problems that limit the competitiveness of QAOA: First, optimizing these parameters requires many calls to the costly quantum hardware, and the use of gradient-based optimizers, which are highly successful in classical machine learning (ML), makes this - due to the Parameter-Shift rule - even more costly \cite{crooks2019gradients}. Second, variational quantum circuits are known to suffer from the barren plateau phenomenon \cite{mcclean2018barren}, which severely limits the size of the circuits that can be successfully trained. So even if one accepts the high cost of parameter optimization (in terms of operational cost and wall-clock time), it is increasingly unlikely to do so successfully as problem sizes and correspondingly quantum circuit size and depth increase.

To mitigate these two problems, many recent studies focus on fixed-parameter variations of QAOA, where the parameters are initialized using a predefined strategy and are fixed instead of being optimized \cite{montanez2025towards}. In this work, we focus on linear ramp QAOA (LR-QAOA) \cite{montanez2025towards}, which uses a linear ramp initialization for the $\gamma$ and $\beta$ parameters of the layers of QAOA, motivated by quantum annealing \cite{apolloni1989quantum}. LR-QAOA reduces the complexity of classical QAOA, where $2p$ parameters need to be optimized, to a problem where only three parameters need to be optimized: the slope of the $\gamma$ and $\beta$ parameters across layers $\Delta_{\gamma}$ and $\Delta_{\beta}$, and the number of layers $p$. Crucially, determining the optimal values for these three parameters for a given problem instance remains an unsolved challenge. Addressing this is the central goal of our work. Put simply, we aim to find a method for defining these three parameters a priori. 

LR-QAOA holds the promise of one-shot optimization provided that the remaining variable parameters can be predetermined. This would substantially reduce the wall-clock time required for the overall optimization procedure, eliminating the need for repeated quantum circuit executions and requiring only a single evaluation of the LR-QAOA circuit. Practical quantum advantage can only be demonstrated through superiority in total wall-clock time, which includes both quantum execution and classical overheads. Consequently, realizing a genuine one-shot QAOA protocol (without reliance on iterative classical parameter optimization or exhaustive empirical searches over parameter sets) would mark a significant milestone towards the efficient applicability of QAOA to real world problems. Furthermore, this one-shot paradigm would lower the resource demands and operational costs associated with deploying the quantum algorithm.

To achieve this, we propose QAOA-Predictor, a model for forecasting LR-QAOA's success probability (defined as the probability of sampling the optimal solution). Our model, which is implemented as a Graph Neural Network (GNN) \cite{scarselli2008graph}, is trained on a large and diverse set of combinatorial optimization problems with varying ramp parameters and layer numbers. The trained model than predicts the appropriate parameter values for unseen problem instances, ensuring the best possible solution within the parameter solution space is found while minimizing computational resource usage. Instead of time consuming parameter optimization or empirical evaluations for each new problem instance - as is typical in standard QAOA - we shift this optimization overhead to the prior training phase of our model. Our method allows for fast, one-shot optimization with LR-QAOA, with one important constraing: Our approach is only able to give optimal parameter configurations within the specified (training) solution space. As has been shown in prior works, empirical evidence suggests that LR-QAOA (and most likely any other version of QAOA) is not a general purpose solver \cite{montanez2025towards}. Therefore, LR-QAOA is not expected to find the optimal solutions of all problem instances with a high or even non-zero probability. This means that QAOA-Predictor can be used in two different ways: Either to identify suitable promising problem instances together with high-quality parameters for the execution of LR-QAOA, or to flag  improper problem instances for which alternative algorithms are likely more effective.

\section{Quantum Computing for Combinatorial Optimization}

There are several quantum algorithms for combinatorial optimization, each with unique advantages and disadvantages. A crucial distinction must be made between algorithms requiring a fault-tolerant quantum computer and those suitable for currently available, so called Noisy-Intermediate-Term-Quantum (NISQ) devices \cite{preskill2018quantum}. Within the fault-tolerant class, two common examples are Grover's algorithm for unstructured search \cite{grover1996fast} and the HHL algorithm for solving linear systems of equations \cite{Harrow_2009}. Grover's algorithm can speed up brute-force searches of the solution space, while the HHL algorithm efficiently solves (some) linear systems of equations frequently encountered in classical optimization.

Within the class of quantum algorithms which can be executed on current NISQ hardware, the two most popular are the Variational Quantum Eigensolver (VQE) \cite{tilly2022variational}, which is typically also used for chemistry-related problems \cite{kandala2017hardware}, and QAOA \cite{farhi2014quantum}, which is primarily applied to combinatorial optimization problems. Since their initial introduction, several improvements to these algorithms have been proposed, such as ADAPT-VQE \cite{grimsley2019adaptive} or Warm-Start QAOA \cite{egger2021warm}.

A mutual feature of these quantum optimization algorithms for classical problems is that they require a problem Hamiltonian: The optimal solution to the problem is typically encoded in the ground state of this Hamiltonian, allowing the optimization task to be defined as the minimization of the Hamiltonian's energy. The most common method for translating a combinatorial optimization problem into this quantum-usable format is the QUBO formulation.

\subsection{Quadratic Unconstrained Binary Optimization}

A QUBO problem is defined as the minimization of an unconstrained quadratic objective function over a set of binary variables. Mathematically, the QUBO problem seeks to find a binary vector $\mathbf{x} = (x_1, x_2, \ldots, x_n)$, where $x_i \in \{0, 1\}$, that minimizes the function $H_Q(\mathbf{x})$:

\begin{equation}
    \min_{\mathbf{x} \in \{0, 1\}^n} H_Q(\mathbf{x}) = \sum_{i=1}^n \sum_{j=i}^n Q_{ij} x_i x_j.
    \label{eq:qubo_sum}
\end{equation}

This formulation can be equivalently expressed in matrix notation as:

\begin{equation}
    \min_{\mathbf{x} \in \{0, 1\}^n} H_Q(\mathbf{x}) = \mathbf{x}^{\mathrm{T}} \mathbf{Q} \mathbf{x}.
    \label{eq:qubo_matrix}
\end{equation}

\begin{figure}
    \centering
    \includegraphics[width=0.8\linewidth]{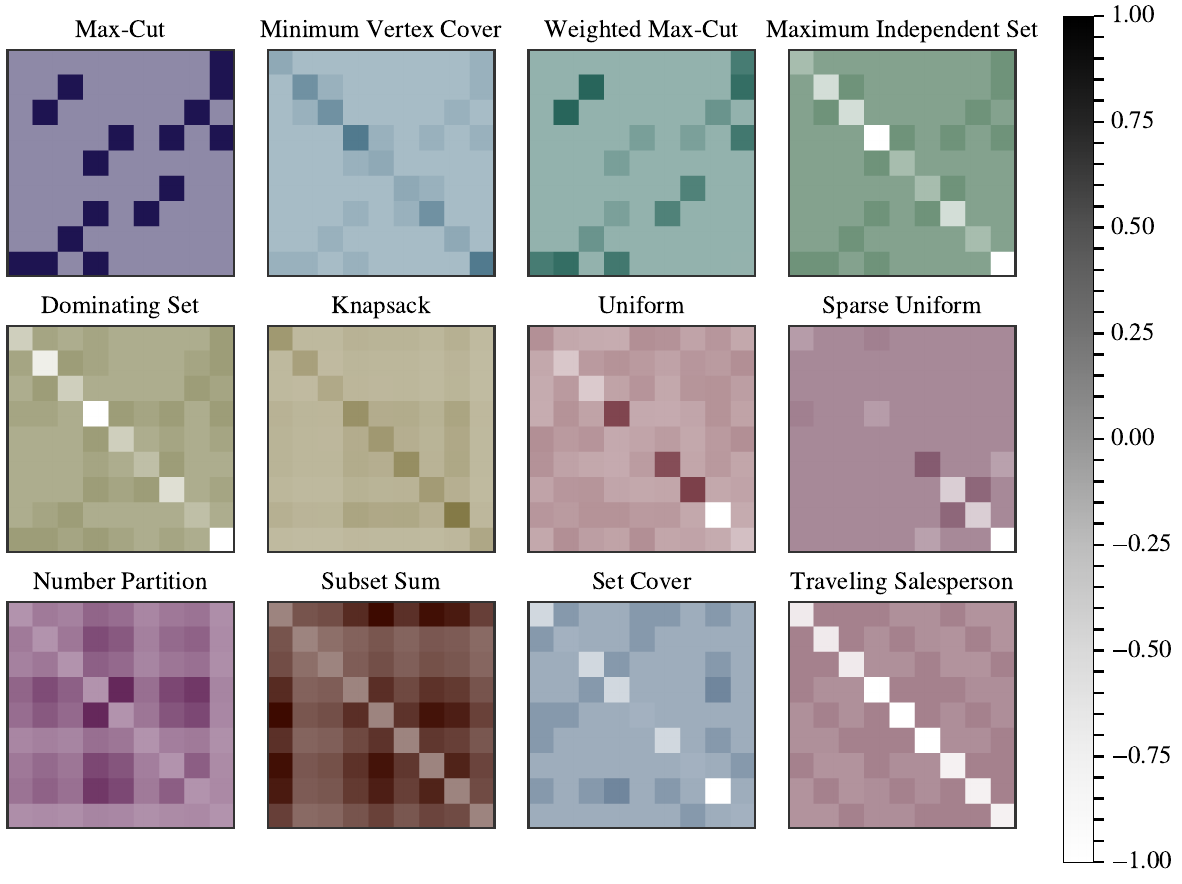}
    \caption{Normalized QUBO matrices of the combinatorial optimization problems considered in this work.}
    \label{fig:qubo_matrices}
\end{figure}

Many challenging NP-hard combinatorial optimization problems can be formulated as QUBO instances. Prominent examples include the Max-Cut, Graph-Coloring and Traveling-Salesperson problems. QUBO instances are, by definition, unconstrained. However, most industrially-relevant problems include constraints, which are typically added to the objective function as penalty terms. Some constraints allow for straightforward, well-known penalty encodings, while others — especially inequality constraints — are significantly harder to formulate and typically require the introduction of auxiliary slack variables. However, alternative methods, which can represent such constraints without relying on additional variables, have recently been developed \cite{montanez2024unbalanced}. For a general overview on how to formulate QUBO problems and encode constraints, the reader is referred to \cite{glover2018tutorial}. Exemplary QUBO matrices of the problem classes considered in this work can be seen in Fig. \ref{fig:qubo_matrices}.

\subsubsection{Translation into Ising Hamiltonian}

A QUBO instance can be easily converted into an Ising Hamiltonian via a variable change, which is required by the aforementioned quantum algorithms. The Ising model is defined by a Hamiltonian $H_{\text{Ising}}(\mathbf{s})$ over a set of spin variables $\mathbf{s} = (s_1, s_2, \ldots, s_n)$, where each spin $s_i \in \{-1, 1\}$:

\begin{equation}
    \min_{\mathbf{s} \in \{-1, 1\}^n} H_{\text{Ising}}(\mathbf{s}) = \sum_{i<j} J_{ij} s_i s_j + \sum_i h_i s_i ,
    \label{eq:ising_min}
\end{equation}

where $J_{ij}$ are the coupling strengths and $h_i$ are the local fields. The transformation between the binary variables ($x_i$) and the spin variables ($s_i$) is established by the linear mapping 

\begin{equation}
    x_i = \frac{1}{2}(1 + s_i).
    \label{eq:mapping}
\end{equation}

Substituting this mapping (Eq. \ref{eq:mapping}) into the QUBO function (Eq. \ref{eq:qubo_sum}) shows that the minimization of $H_Q(\mathbf{x})$ is equivalent to the minimization of $H_{\text{Ising}}(\mathbf{s})$.

\subsection{QAOA}

QAOA, introduced by Farhi, Goldstone, and Gutmann in 2014 \cite{farhi2014quantum}, is a hybrid quantum-classical variational algorithm designed to find approximate solutions to combinatorial optimization problems, particularly those formulated as QUBO instances that have been introduced in the previous section. 

QAOA operates by preparing a parameterized quantum state $|\psi(\boldsymbol{\gamma}, \boldsymbol{\beta})\rangle$ on $n$ qubits. The goal is to minimize the expectation value of the cost (or problem) Hamiltonian $H_C$, which encodes the objective function to be minimized.

For a QUBO problem $\mathbf{x}^\top \mathbf{Q} \mathbf{x}$, the corresponding classical cost Hamiltonian is the Ising-like operator

\begin{equation}
    H_C = \sum_{i \leq j} Q_{ij} Z_i Z_j,
    \label{eq:hc_qubo}
\end{equation}

where $Z_i$ is the Pauli-Z operator on qubit $i$, and the diagonal linear terms are included when $i=j$.

The QAOA ansatz with the number of layers $p$ is constructed as follows: First, a uniform superposition state (the “mixing” ground state) is prepared as:
   \[
   |+\rangle^{\otimes n} = H^{\otimes n} |0\rangle^{\otimes n},
   \]
   where $H$ is the Hadamard gate.

Then, $p$ layers are applied of cost-unitary evolution $U_C(\gamma_k) = e^{-i \gamma_k H_C}$ and mixing-unitary evolution $U_M(\beta_k) = e^{-i \beta_k H_M}$ operators, with $H_M = \sum_{i=1}^n X_i$ (the transverse-field mixer):

\[
|\psi(\boldsymbol{\gamma}, \boldsymbol{\beta})\rangle = U_M(\beta_p) U_C(\gamma_p) \cdots U_B(\beta_1) U_C(\gamma_1) \, |+\rangle^{\otimes n}.
\]

During the optimization, the variational objective is then
   
\[
\min_{\boldsymbol{\gamma}, \boldsymbol{\beta}} \langle \psi(\boldsymbol{\gamma}, \boldsymbol{\beta}) | H_C | \psi(\boldsymbol{\gamma}, \boldsymbol{\beta}) \rangle.
\]
   
This expectation value is estimated on a quantum device, and typically a classical optimizer (e.g., COBYLA \cite{Powell1994}, SPSA \cite{spall1987stochastic}, or gradient descent) updates the $2p$ parameters of the alternating layers iteratively, while the number of layers $p$ is chosen empirically.

As $p \to \infty$, QAOA recovers the exact quantum adiabatic algorithm with a linear schedule, and for certain problems (e.g., Max-Cut on bounded-degree graphs) it can achieve approximation ratios that improve with $p$ \cite{wang2018quantum}. For finite $p$, QAOA can be executed on current quantum devices, at least for low to moderate $p$ ($ 1 \leq p \leq 20$ in current NISQ experiments \cite{pelofske2023high}).

\subsection{LR-QAOA}

While, in theory, increasing $p$ improves the performance of QAOA, the practical performance often plateaus or even degrades beyond moderate depths due to barren plateaus in the optimization landscape and the increasing difficulty of training the parameters with local optimizers \cite{larocca2025barren, gleissner2024restricted}. Additionally, on current hardware, deeper circuits amplify error accumulation, making effective parameter optimization without error mitigation or correction extremely difficult \cite{xue2021effects}. Several strategies have been proposed to mitigate these issues while preserving or improving solution quality as can be seen in \cite{montanez2025towards,vcepaite2025quantum}

\begin{figure}[tbp]
    \centering
    \includegraphics[width=0.48\linewidth]{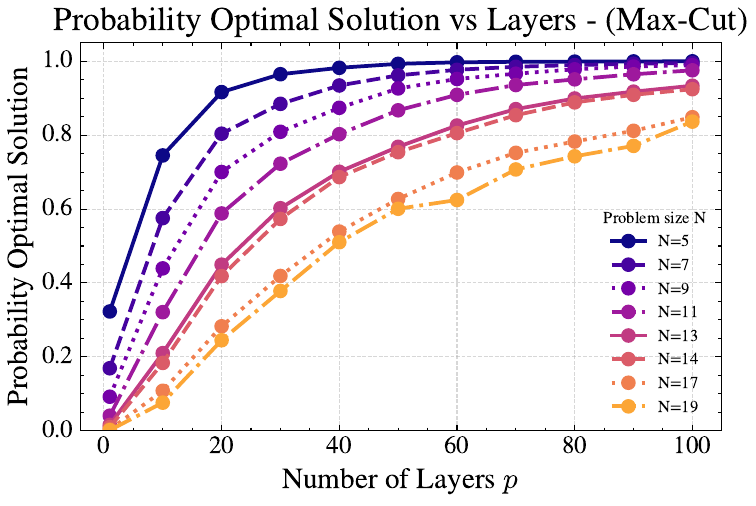}
    \hfill
    \includegraphics[width=0.48\linewidth]{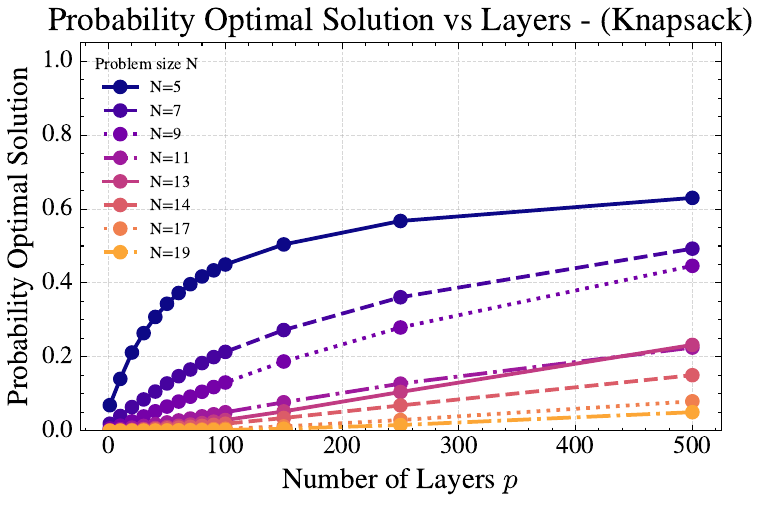}
    \caption{Performance of LR-QAOA with fixed $\Delta_{\gamma, \beta}=0.3$ parameter on two exemplary problem classes. For each problem class (Max-Cut and Knapsack) and size $N \in [5, 19]$, 150 random instances are sampled and the mean probability of the optimal solution across these samples is plotted for increasing number of layers $p$. For the Max-Cut problem class (left), a problem size of $N=7$ requires $p=20$ layers for a $80\%$  probability of sampling the optimal solution, while a larger instance of $N=14$ requires $p=60$ layers to achieve the same probability. For the Knapsack problem class (right), the effect of compensating size via layers is not as strongly pronounced. Here the overall probability is not only lower, but also more layers are required to increase the probability of the optimal solutions.}
    \label{fig:mean_trend_dataset}
\end{figure}

A particularly effective hardware-efficient variant of QAOA is LR-QAOA as developed in recent works by Montañez-Barrera et al. \cite{montanez2025towards}. Instead of optimizing $2p$ independent parameters, LR-QAOA employs fixed linear ramp schedules for the parameters $\gamma_k$ and $\beta_k$, eliminating the need for classical optimization (see Eq. \ref{eq:schedule}). This deterministic, non-variational approach yields a universal set of parameters that is able to approximate optimal solutions across diverse combinatorial optimization problems, with performance generally improving with increasing $p$ (see Fig. \ref{fig:mean_trend_dataset}). However, LR-QAOA is not a general purpose solver and does not provide high quality solution across all problem classes (as can be seen in Fig. \ref{fig:mean_trend_dataset} its performance can greatly vary and potentially not scale).

The LR-QAOA ansatz follows the same layered structure as standard QAOA but with parameters (which are not optimized) prescribed by linear ramps:

\begin{equation}\label{eq:schedule}
    \gamma_k = \Delta_\gamma \cdot \frac{k}{p}, \quad \beta_k = \Delta_\beta \cdot \left(1 - \frac{k}{p}\right) \quad \text{for} \quad k = 1, \dots, p,
\end{equation}

where $\Delta_\gamma$ and $\Delta_\beta$ are fixed total ramp amplitudes (often chosen empirically as $\Delta_\gamma = \Delta_\beta = 0.5$, tuned for broad applicability across problem classes \cite{montanez2025towards}). The resulting circuit is scalable in width ($n$ qubits) and depth ($p$ layers) and does not require iterative feedback. This drastically reduces both the overall wall-clock time and the overall computational cost of the algorithm. 

For many problem classes, the probability of finding the optimal solution can be substantially improved by increasing the number of layers $p$ (see Fig. \ref{fig:mean_trend_dataset}). However, this benefit does not apply universally across all problem classes and can be significantly reduced or even absent in certain cases. The relationship between the specific problem class, the size of the instance, the number of layers and the chosen $\Delta_{\gamma, \beta}$ value remains poorly understood, and most parameters are typically selected empirically.

\section{QAOA-Predictor}\label{sec:qaoa-predictor}

Despite its considerable advantages, LR-QAOA faces significant challenges. Even though the ramp parameter $\Delta_{\gamma, \beta}$ and the number of layers are not themselves optimized, appropriate values still need to be determined empirically. Consequently, the probability of sampling the optimal solution for a new problem instance is unknown when executing LR-QAOA with arbitrary parameters. In real world scenarios, this is highly problematic, since the quality of the given solution cannot be inferred. This is contrary e.g. to the Goemans-Williamson algorithm for solving Max-Cut which gives a lower bound for the approximation ratio of the solution \cite{goemans1995improved}. An important step towards the applicability of LR-QAOA to real world problems would therefore be prior knowledge of the behavior of the algorithm for a given problem instance.

Several previous studies have addressed the challenge of parameter optimization through transfer learning — the strategy of reusing $\gamma,\beta$ parameters optimized on one instance for different problem instances. For example, \cite{montanez2025transfer} introduced a framework where parameters optimized on small instances are applied to larger instances or even different problem classes. Their findings highlight that the Bin Packing Problem generates the most transferable parameters, potentially serving as a universal starting point for optimization. Similarly, \cite{lyngfelt2025symmetry} investigated the transferability of parameters between Max-Cut instances based on symmetry and graph properties. Their research suggests that while transferred parameters can provide effective warm-start guesses, structural differences between graphs often limit their success. For instance, they observed that transferring parameters between graphs with different parities generally results in poor performance.

In contrast to transfer learning strategies, we propose QAOA-Predictor, is a predictive framework designed to forecast the success probability of LR-QAOA — that is, the likelihood of sampling the optimal solution. Rather than directly optimizing the $\gamma, \beta$ parameters, our approach predicts the success probability for a specific problem instance and parameter set of ramp parameter and layer number. QAOA-Predictor serves as a guide for selecting the most effective parameters before running the algorithm, thereby avoiding expensive runtime optimization or extensive empirical tuning. A step-by-step overview of our method is shown in Fig.\ref{fig:method}. It shows both the training routine and the following usage of QAOA-Predictor. Each instances' Ising Hamiltonian matrix is combined with the value of $\Delta_{\gamma,\beta}$ and $p$ for each prediction. Ultimately, for a given problem instance, the model enables identification of the minimal number of layers required to achieve a desired sampling probability of the optimal solution.

Our contribution is twofold. First, by moving the parameter selection overhead to a one-time training phase, QAOA-Predictor enables rapid, resource-efficient deployment of LR-QAOA, greatly reducing its computational costs during runtime. Second, our method enables the fast identification of suitable (and unsuitbale) use cases for LR-QAOA. Instead of applying a general-purpose quantum algorithm to every problem, QAOA-Predictor can predetermine if LR-QAOA is appropriate for a specific instance without requiring execution on real quantum hardware or through intensive classical simulation. Accordingly, QAOA-Predictor can be employed in two complementary ways: to select promising instances together with high-quality parameters for effective LR-QAOA execution, or to flag difficult instances for which finding optimal solutions is unlikely and should hence not be solved using LR-QAOA. This enables the following optimization workflow: A user begins with an optimization problem and converts it into a QUBO formulation. They then specify the desired minimum probability of sampling the optimal solution along with the maximum number of QAOA layers feasible (dependent on the currently available quantum hardware). The QAOA-Predictor can determine in milliseconds whether this target is achievable. If the prediction is positive, the user can adopt the recommended parameters and the smallest number of layers that still maintain the desired probability, obtaining the optimal solution with a single call to the quantum hardware. \\

In this work, we propose to use a GNN as predictor model for LR-QAOA. We benchmark this model against another deep learning model, a CNN, as well as a ML approach, the k-nearest neighbors (KNN) method and demonstrate its superiority over the other two approaches on a variety of validation tests, created to highlight important features of QAOA-Predictor. First, the main requirement for the trained model is the (approximately) correct prediction on samples that are generated from the same distribution used to generate the training data - \emph{interpolation}. Second, it needs to be able to \emph{extrapolate} to different scenarios: a) larger problem sizes, b) larger layer numbers and c) unseen problem classes. Finally, it should be possible to finetune the trained model using sparse and costly data to improve its predictions on specific instances.

\subsection{Data}\label{sec:Data}

Training a ML model to predict the performance of LR-QAOA for a large range of combinatorial optimization problems requires an extensive and diverse dataset. This dataset should contain various problem classes as well as a high number of instances within the same class to ensure the model generalizes well to unseen problems and instances.

The problem classes considered in this work can be split into two types: graph-based and non-graph based. In total, we consider 12 problem classes (see Fig. \ref{fig:qubo_matrices} and Fig. \ref{fig:data_embedding}), most of which, such as Max-Cut and Knapsack, have been thoroughly investigated and typically exhibit a consistent structure. Additionally, we also consider two artificially generated problem classes where the coefficients of the QUBO matrix are uniformly sampled in either a dense or sparse manner (Uniform and Sparse Uniform, respectively) to evaluate whether the model can learn to predict the performance of LR-QAOA on uncommon QUBO structures.

To train a model using supervised learning, the dataset must consist of input features paired with ground-truth labels. While our input data format varies depending on the predictor model (GNN, CNN or KNN), the target labels remain consistent across models. The data generation process begins by randomly generating a QUBO matrix for a certain problem class, which is converted into an Ising Hamiltonian using the mapping in Eq. \ref{eq:mapping}. Following the methodology in \cite{montanez2025towards}, we normalize the Hamiltonian by its maximum quadratic coefficient. This normalized Hamiltonian is then stored in two formats: as a raw matrix for the CNN and as a graph representation for the GNN. For the KNN approach, we compute a set of seven matrix features $F$ which can be easily and quickly computed (see Tab. \ref{tab:matrix_features}).

\begin{table}[htbp]
\centering
\caption{Structural matrix features derived from the Ising Hamiltonian matrices.}
\label{tab:matrix_features}
\begin{tabular}{ll}
\toprule
\textbf{Feature}         & \textbf{Description} \\
\midrule
density                  & Edge density $= m / \binom{n}{2}$ \\
num. edges               & Raw number of edges $m$ \\
avg. degree              & Average node degree $= 2m/n$ \\
mean diag.               & Mean of absolute diagonal entries $\frac{1}{n}\sum_i |A_{ii}|$ \\
std. diag                & Standard deviation of absolute diagonal entries \\
mean off-diag.          & Mean of absolute off-diagonal entries $\frac{1}{n(n-1)}\sum_{i\neq j} |A_{ij}|$ \\
std. off-diag.          & Standard deviation of absolute off-diagonal entries \\
\bottomrule
\end{tabular}
\end{table}

Since our goal is to predict the performance of LR-QAOA based on the number of layers $p$ and the ramp parameter $\Delta_{\gamma,\beta}$, we execute the algorithm under these varying conditions for each instance. The resulting label is the probability of obtaining an optimal solution - a bit string that yields the minimal cost - for the respective parameter configuration. Our models are trained to predict these ground-truth probabilities calculated during the data generation phase. The probabilities of the optimal solutions of the LR-QAOA circuit are computed with a pennylane statevector simulator \cite{bergholm2018pennylane} with 10000 shots. However, the data generation (as will be discussed in Sec. \ref{sec:finetuning}) can also be performed on a quantum device. To calculate the true optimal values of the problem instances, we perform a brute-force search across all $2^n$ bit strings to identify the global optima, then sum their respective probabilities within the LR-QAOA output distribution for the resulting ground-truth probabilities. This procedure is repeated for multiple instances $I$ for all  (non-)graph-based problems described above with varying sizes $N$ to construct the full datasets.

For each dataset, we generate an equal amount of graph-based and non-graph based instances. All instances are randomly generated to ensure variability and diversity between different instances of the same problem class. For example, when a Max-Cut-Weighted instance is generated, a graph-type, such as Erdos-Renyi, is sampled alongside a certain probability of generating an edge between any two nodes. Then, the weights of the edges are also uniformly sampled between a fixed interval. For a more complete overview of the different problem classes and datasets, the reader is referred to the Appendix \ref{secA1}.

We generate three different datasets, for which the number of problem classes remains constant. The validation dataset as well as the finetuning dataset contain instance sizes and layer numbers beyond the original training dataset in order to evaluate the extrapolation capabilities of the models:

\begin{itemize}
    \item \textbf{Training Dataset:} Contains $I=50$ instances for each problem class with $N=[5,15]$ and $p\in\{1,10,20,30,40,50,60,70,80,90,100\}$. Used for training the models.
    \item \textbf{Validation Dataset:} Contains $I=10$ instances for each problem class with $N=[5,20]$ and $p\in\{1,10,20,30,40,50,60,70,80,90,100,250,500\}$. Used for validating the trained model's performance for unseen instances and for extrapolating to larger numbers of layers as well as for larger problem instances.
    \item \textbf{Finetuning Dataset:} Contains $I=1$ instances for each problem class with $N=[5,20]$ and $p\in\{1,10,20,30,40,50,60,70,80,90,100,250,500\}$. Used for finetuning the trained model on additional instances to enhance the extrapolation to larger numbers of layers as well as to larger problem instances.
\end{itemize}

\subsection{Training Pipeline}

\begin{figure}[tbp]
    \centering
    \includegraphics[width=0.85\linewidth]{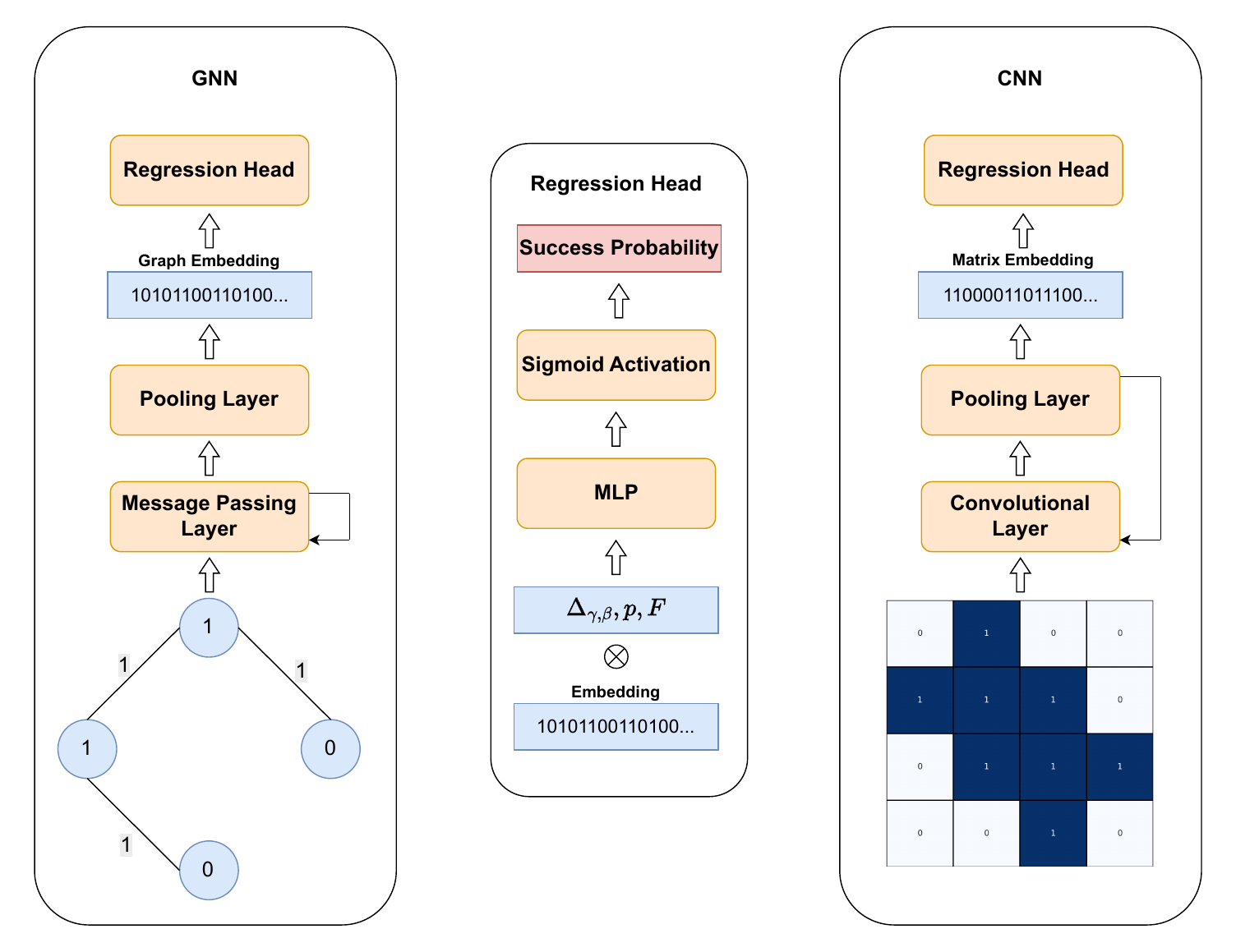}
    \caption{The GNN and CNN models share a similar overall structure, differing only in their input format and initial processing layers. The GNN uses a graph-based backbone to process problem instances as graphs, while the CNN uses a convolutional backbone to process them as matrices.After this initial feature extraction, both models use the same regression head. This head concatenates the backbone's embedding with the parameter $\Delta_{\gamma,\beta}$, and the layer number $p$. This combined vector is then passed through a MLP to predict the final success probability.}
    \label{fig:models}
\end{figure}

With the dataset established, the training procedure of the deep learning models follows a supervised regression framework, where both the GNN-based and CNN-based models share a common objective. Each model is designed to receive a representation of the Ising Hamiltonian - either as a graph or a matrix - the specific number of layers $p$ and the ramp parameter $\Delta_{\gamma,\beta}$. Based on these inputs, the models predict the success probability of LR-QAOA for that instance, and the absolute error between the prediction and the previously simulated true probability is computed.

\paragraph{GNN}
The GNN-based model interprets the Ising Hamiltonian as a weighted graph, where nodes represent variables and edges represent the quadratic interactions. The GNN used in this work follows the Deep Sets architecture \cite{zaheer2017deep}, which guarantees permutation invariance and ensures that the graph representation is independent of node ordering.  The model begins by performing message-passing operations to generate N distinct node embeddings that capture the local structure of the optimization problem (see Fig. \ref{fig:models}). To transition from these local features to a global representation, the model applies both global mean and max pooling to the node embeddings. The results of these pooling operations are concatenated to form a unique graph embedding. This embedding is then concatenated with the scalar inputs $p$ and $\Delta_{\gamma,\beta}$ and passed to a final Multi-Layer Perceptron (MLP). The MLP maps these combined features to a single output value, which is bounded between $[0,1]$ via a Sigmoid activation function to represent the predicted success probability.

\paragraph{CNN} 
The CNN-based model follows a similar structure but utilizes a convolutional backbone to extract features directly from the Ising Hamiltonian matrix (see Fig. \ref{fig:models}). In this approach, the matrix is treated as a single-channel input, and a series of convolutional layers and pooling operations are used to identify spatial correlations within the quadratic coefficients. The resulting feature map is flattened into a one-dimensional embedding of the matrix. Following this extraction, the model utilizes the same MLP structure as the GNN model, concatenating the matrix embedding with $p$ and $\Delta_{\gamma,\beta}$ before applying a final Sigmoid activation. By maintaining an identical regression head for both architectures, any variations in performance can be attributed to the effectiveness of the respective graph or matrix encoding strategies, as well as to the graph and convolutional backbones. 

\paragraph{KNN} 
We use a simple distance-weighted KNN with $k=10$ neighbors, which is trained and evaluated on the same training and validation dataset as the deep learning models. The KNN used in this work serves as a baseline to evaluate the performance of the deep learning models. We therefore do not perform a large scale hyperparameter search, but rather use typical values in this work. Details of the used KNN parameters can be found in the Appendix (see Tab. \ref{tab:knn-params}).

\section{Results}\label{sec:results}

Our evaluation begins with a preliminary study using Principal Component Analysis (PCA) to examine the distribution of basic features across the different problem classes to analyze the structure and diversity of the generated dataset. Following this analysis, we test the capabilities of the three models with a focus on two critical aspects: generalization to unseen instances generated using the same distribution as the training dataset and extrapolation to unseen instances which diverge in some key feature(s) from the training dataset. Specifically, we assess the models' ability to maintain accuracy when applied to larger instance sizes $N$ and higher numbers of layers $p$ (than those encountered during the training phase). We also test the models ability to transfer its learned representations to instances from unseen problem classes. This evaluation allows us to determine the extent to which the models have learned the underlying landscape of LR-QAOA performance rather than merely memorizing specific problem class configurations.

\subsection{Preliminary Analysis}

Before training the predictor models, we perform a simple PCA analysis of the dataset using the matrix features derived from the Ising Hamiltonian's matrices discussed above (see Tab. \ref{tab:matrix_features}). The dataset must contain structure which the models can use to predict the performance of LR-QAOA. Previous results by Montañez-Barrera et al. \cite{montanez2025towards} show that the choice of ramp parameter $\Delta_{\gamma,\beta}$ as well as the overall performance of LR-QAOA greatly deviates between problem classes such as Max-Cut and Knapsack  (also see Fig. \ref{fig:mean_trend_dataset}). Hence, strong separation of these problem classes in the PCA-projected feature space would be advantageous.

\begin{figure}
    \centering
    \includegraphics[width=1.0\linewidth]{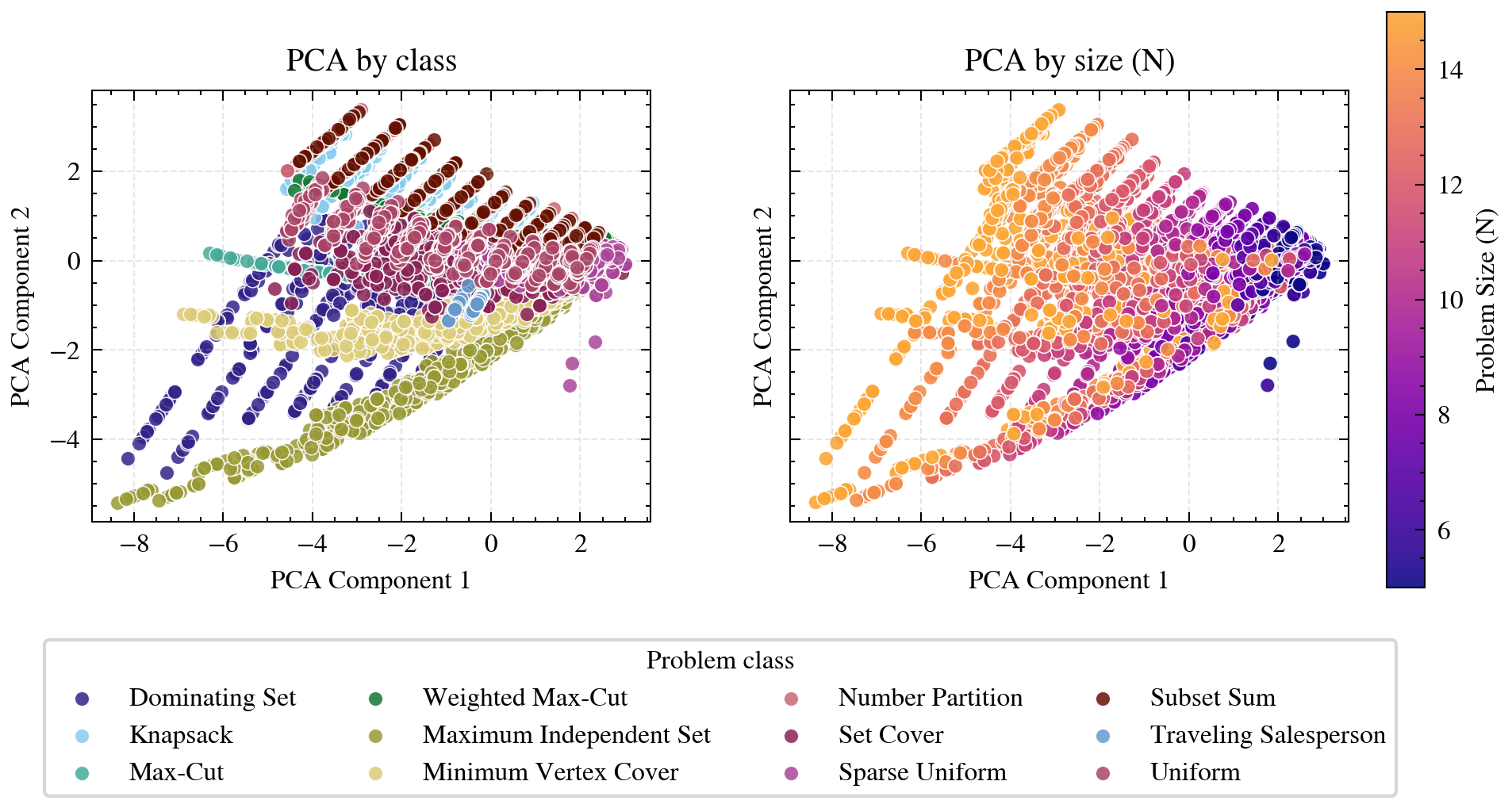}
    \caption{Embedding of the validation data set using PCA colored by class (left) and size (right)}
    \label{fig:data_embedding}
\end{figure}

The result of the PCA analysis, illustrated in Fig. \ref{fig:data_embedding}, reveals that most of the problem classes can be effectively separated within the two-dimensional space defined by the first two principal components. Problem classes like the Maximum Independent Set or the Minimum Vertex Cover occupy distinct regions in this simple feature space. Problem classes - such as the Dominating Set or the Subset Sum - even show clear differences between instances of varying size $N$. This clear clustering demonstrates that the instances and their respective problem classes possess distinct structural characteristics that are distinguishable from each other. In fact, this can already be seen in Fig. \ref{fig:qubo_matrices}, where most of the problem classes show clear and varying structures. This separation suggests that the feature space is sufficiently rich for a ML model to identify patterns, which likely correlates with the models' capacity to learn and predict the performance of LR-QAOA across different problem classes. 

\subsection{Hyperparameter Tuning of the GNN and CNN models}\label{sec:tuning}

To ensure a fair comparison, we conducted a thorough hyperparameter tuning process for both deep learning models. We use a grid-search method to test every combination of the selected hyperparameters. The complete list of these parameters and their values is provided in Appendix \ref{secA:hyperparameters}.

\begin{figure}
    \centering
    \includegraphics[width=0.49\linewidth]{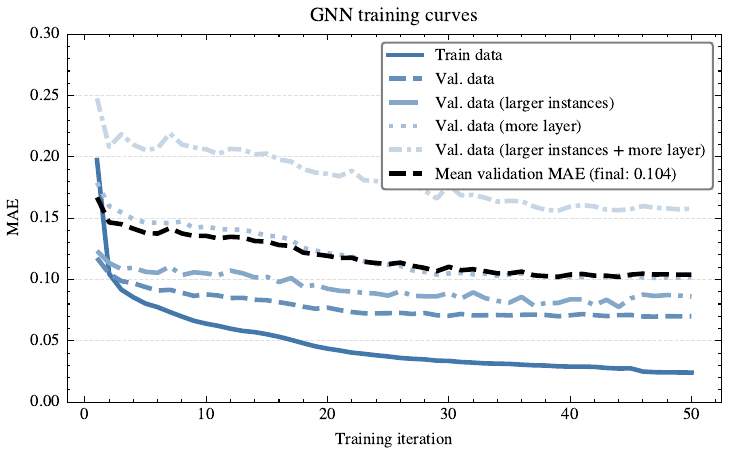}
    \hfill
    \includegraphics[width=0.49\linewidth]{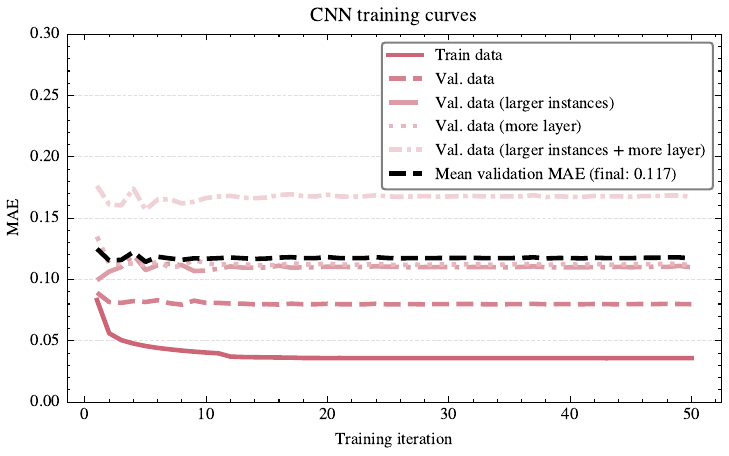}
    \caption{Training curves of both GNN and CNN models for the best set of hyperparameters found.}
    \label{fig:training}
\end{figure}

Fig. \ref{fig:training} shows the training and four validation losses of the best found hyperparameter sets for both the CNN and GNN models. Next to the standard validation loss used to identify the interpolation capability of the model, three additional loss terms are used. These terms, discussed in more detail in Secs. \ref{sec:exp_size} and \ref{sec:exp_layer}, monitor the models' extrapolation capabilities during training. The mean of all validation loss terms serves as the criterion for the best set of hyperparameters. As can be seen, the overall GNN outperforms the CNN in both, training and mean validation loss. In contrast to the continuous decrease in validation loss observed for the GNN throughout training, the CNN exhibits early stagnation and shows no further reduction after the fifth epoch.

\subsection{Predictions of LR-QAOA}

\begin{figure}
    \centering
  \includegraphics[width=0.45\linewidth]{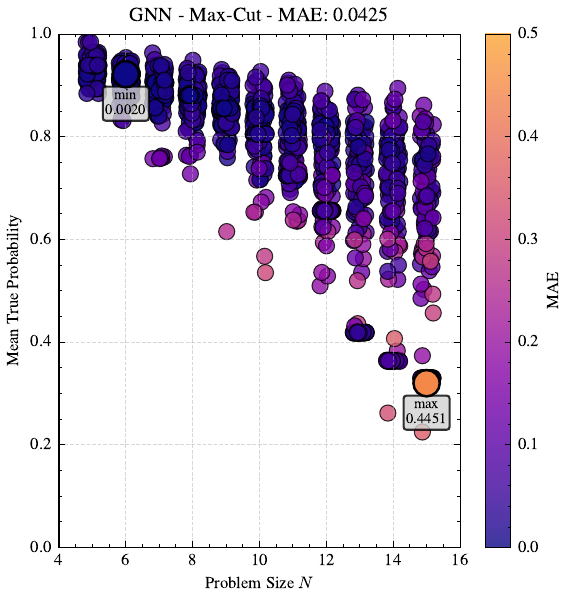}
    \hfill
    \includegraphics[width=0.45\linewidth]{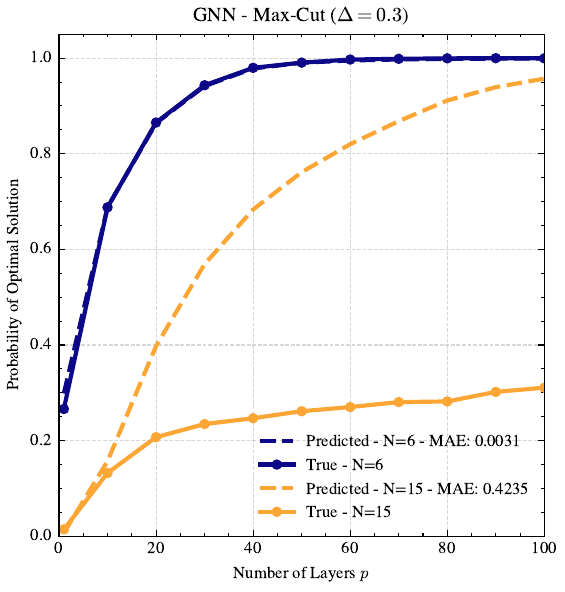}
  \vspace{8pt}
  \includegraphics[width=0.45\linewidth]{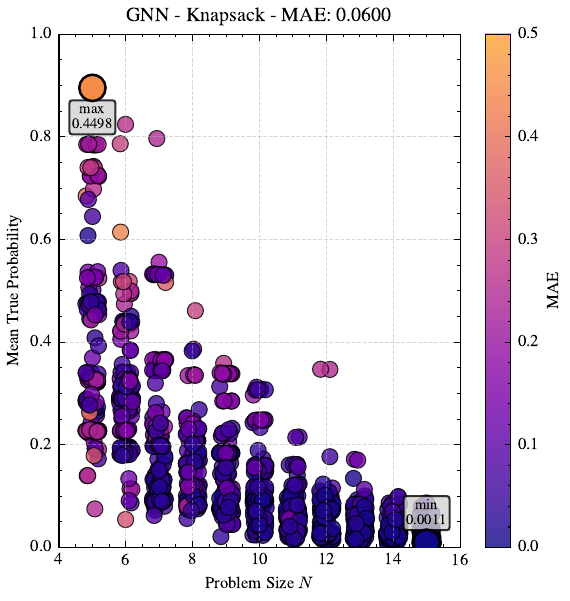}
    \hfill
    \includegraphics[width=0.45\linewidth]{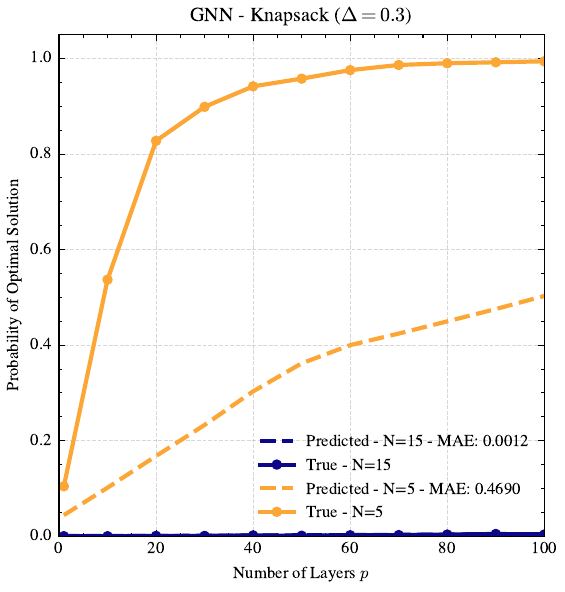}
    \caption{Prediction of GNN on the problem classes Max-Cut (top) and Knapsack (bottom) of the validation dataset. For both classes the resulting MAE on the validation dataset for layers 1 to 100 is shown in the scatter plots (left). The best and the worst prediction for both classes based on the MAE of the instances across all layers is shown in the line plots, colored by the MAE (right).}
    \label{fig:prediction}
\end{figure}

The trained models predict the probabilities of the optimal solution for a given LR-QAOA instance. As has been discussed in Sec. \ref{sec:qaoa-predictor}, each instances' Ising Hamiltonian matrix is combined with the values of $\Delta_{\gamma,\beta}$ and $p$ for each prediction. Ultimately, for a given problem instance, we are interested in the minimal amount of layers required to sample with LR-QAOA the optimal solution with a certain probability. We are therefore interested in the overall mean absolute error (MAE) across all predicted layers $p$ and $\Delta_{\gamma,\beta}$ for a given problem instance. This can be seen in Fig. \ref{fig:prediction}, where each point in the scatter plot shows the mean true optimal probability across all sampled layers $p = \{1, 10, ... 100 \}$ and parameter $\Delta_{\gamma,\beta}$ and is colored by the MAE of the models prediction for all these instances. For the two well known problems Max-Cut and Knapsack, we marked the instances of the validation dataset where the error of the prediction of the GNN model is highest and lowest and plot their true and predicted probabilities for $\Delta_{\gamma,\beta}=0.3$ to illustrate the models best and worst behavior: Generally speaking, both problem classes show a clear trend for most of the sampled instances as well as some outliers, where the performance of LR-QAOA is \textit{"atypically"} good (Knapsack) or bad (Max-Cut). For \textit{"typical"} problem instances, the predictions of the trained model show a low MAE and recover the behavior of LR-QAOA accurately. This is the case for instances where the increase of layers greatly enhances the optimal probability (Max-Cut) as well as for instances where the probability is near zero, independent of the number of layers (Knapsack). However, the GNN model fails to predict \textit{"atypical"} outliers and greatly over- or underestimates the performance of LR-QAOA in these cases (for all problem classes see Appendix \ref{secA:dataset}).

\begin{figure}[tbp]
    \centering
    \includegraphics[width=0.49\linewidth]{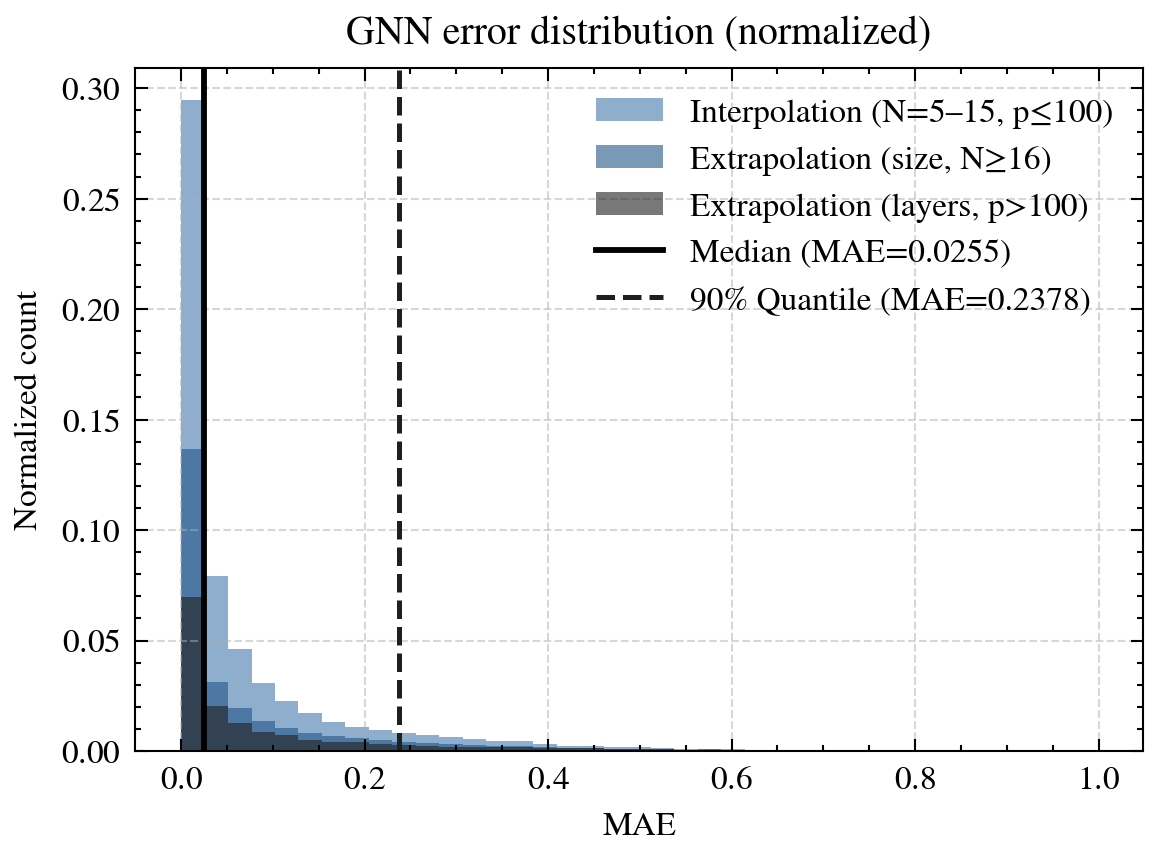}
    \hfill
    \includegraphics[width=0.49\linewidth]{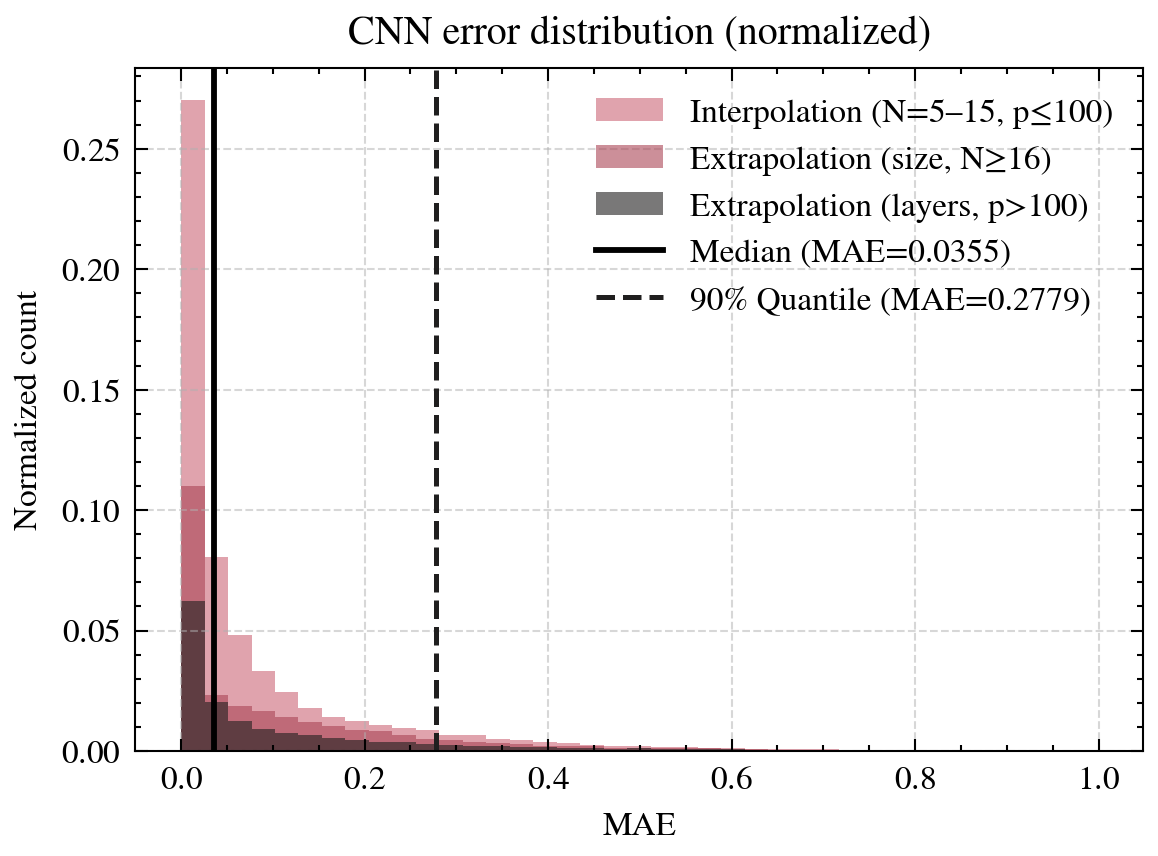}
    \caption{Error distribution of the deep learning models - GNN (left) and CNN (right) - over the validation dataset.}
    \label{fig:histogram_plots}
\end{figure}

A more detailed analysis of the error distribution on the validation dataset shows that the GNN predicts $50\% $ of the samples with an error below $3\%$ and $90 \%$ of the samples with an error below $24\%$ (see Fig. \ref{fig:histogram_plots}). Similarly to the overall training and validation loss, the error distribution of the GNN is lower than that of the CNN, which predicts $90\%$ of the samples with an error below $28\%$, $4\%$ higher than the GNN.

\subsection{Extrapolation to larger problem sizes}\label{sec:exp_size}

\begin{figure}
    \centering
    \includegraphics[width=0.49\linewidth]{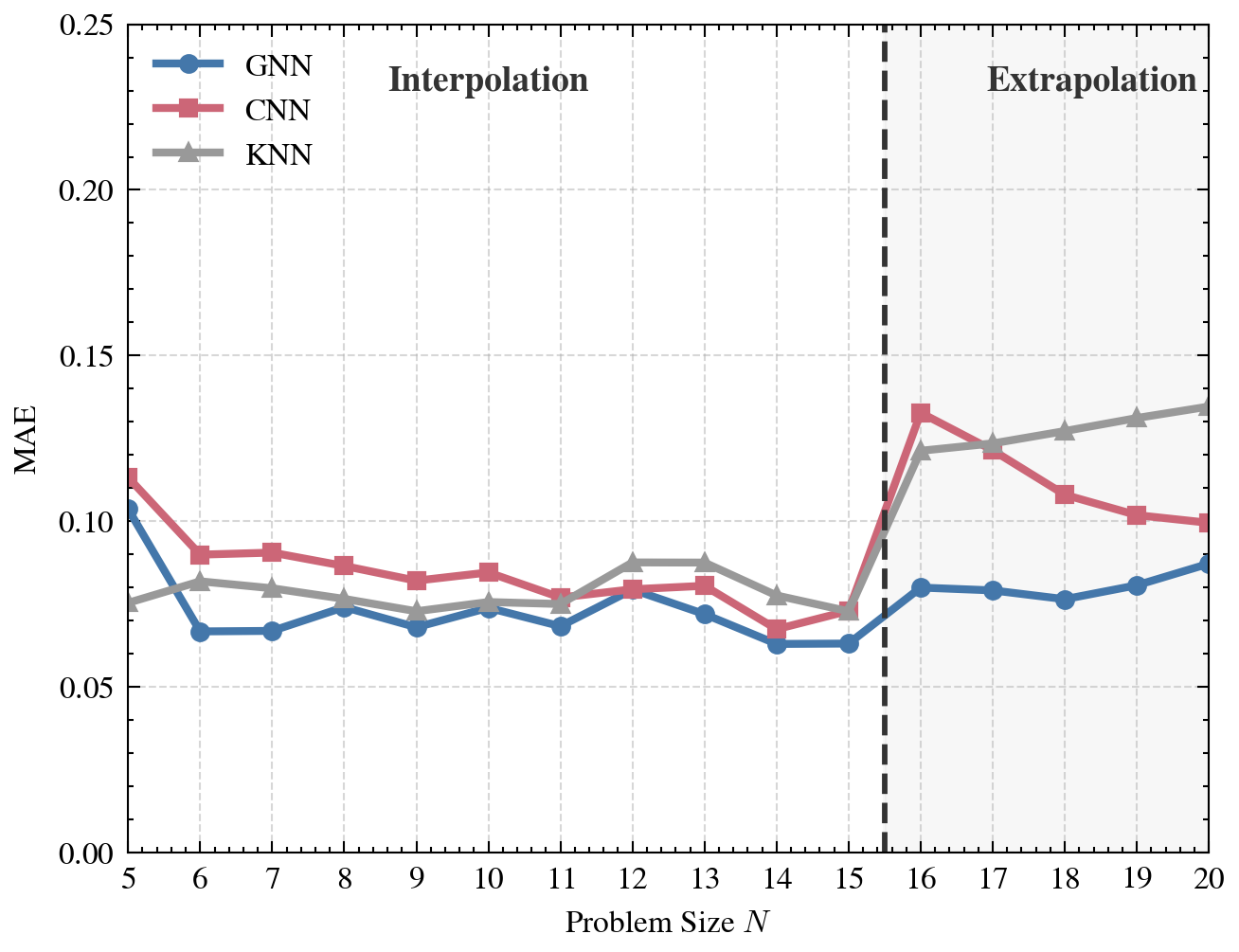}
    \hfill
    \includegraphics[width=0.49\linewidth]{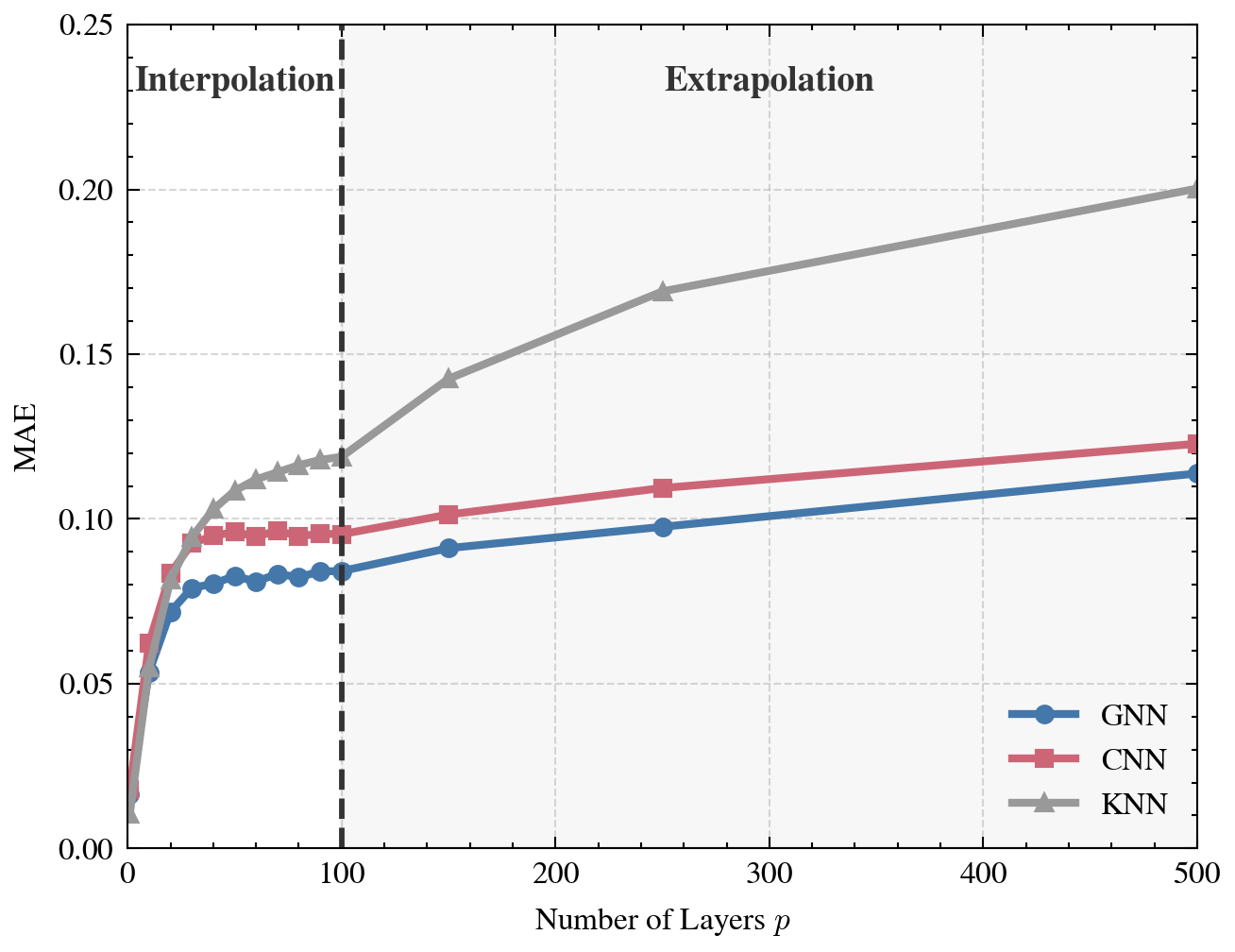}
    \caption{Interpolation and extrapolation of the GNN, CNN and KNN models for different problem sizes $N$ (left) and layer numbers $p$ (right).}
    \label{fig:extrapolation_sizes}
\end{figure}

The most important feature of our prediction model is its strong extrapolation capability toward larger-scale problem instances. This characteristic is of critical importance, given that we aim to provide effective parameter predictions especially for problem sizes where direct optimization rapidly becomes prohibitively expensive or infeasible. Training the model on small, easily solvable instances while achieving reliable performance on significantly larger and much harder instances is a key advantage of QAOA-Predictor. Fig. \ref{fig:extrapolation_sizes} displays the performance of the trained models when interpolating and extrapolating across various instance sizes $N$: the models were trained on sizes $N\in[5,15]$ and tested on the same sizes or larger sizes $N\in[16,20]$. The GNN consistently outperforms the other two models in terms of accuracy except for $N=5$). Notably, the MAE in the extrapolation regions remains comparable low to the interpolation region for the GNN, while the other two models show sharp jumps in MAE. This illustrates that our approach scales well to larger problem instances.

\subsection{Extrapolation to larger numbers of layers}\label{sec:exp_layer}

Another important feature of our prediction model is its ability to extrapolate to larger number of layers. The motivation here is similar to the extrapolation to larger instances: Instances with small $p$ are easy to compute while instances with larger $p$ are increasingly expensive. Fig. \ref{fig:extrapolation_sizes} illustrates how the three trained models perform when interpolating and extrapolating across different numbers of layers $p$. The models were trained with $p=[1,100]$ and are being tested significantly higher $p=\{150,250,500\}$. Generally, the GNN outperforms the CNN and KNN, maintaining a consistently lower MAE for all values of $p$. The accuracy of two deep learning models tends to decrease as $p$ increases. However, for moderate values of $p$ in the extrapolation range, both models maintain a level of performance that is notably similar to their performance during interpolation. These findings underscore the practical robustness of the proposed GNN-based predictor even beyond the training regime, making it a reliable tool for parameter prediction in deeper LR-QAOA circuits.

\subsection{Out-of-Distribution Generalization}

\begin{figure}[tbp]
  \centering
  \includegraphics[width=0.65\linewidth]{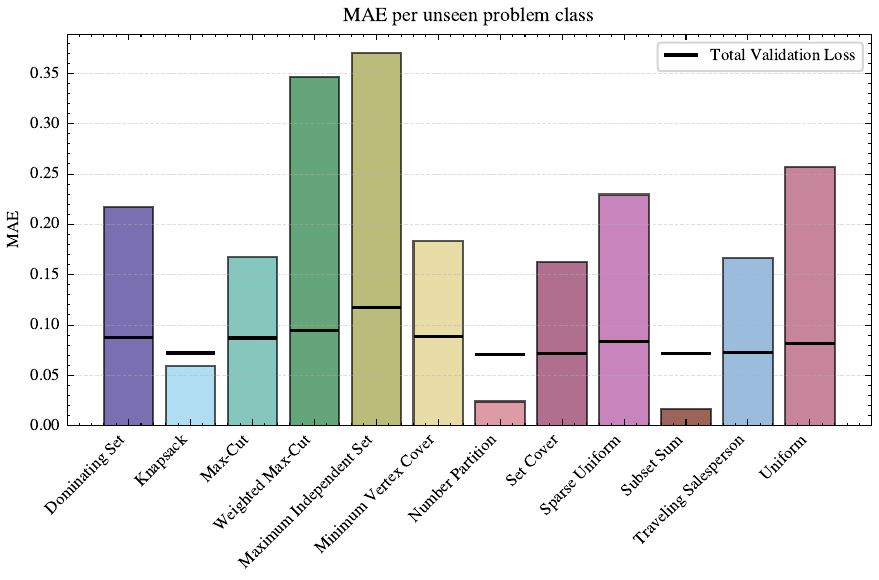}

  \caption{QAOA-Predictor's generalization over problem classes. A GNN-based model is trained on all but one of the problem classes (for all problem classes) and then the mean MAE over the validation dataset is shown for the unseen problem class.}
  \label{fig:out-of-distribution}
\end{figure}

Our third evaluation examines the model's ability to generalize to entirely unseen problem classes. By systematically excluding one entire problem class from the training set, we create a realistic out-of-distribution scenario that mirrors real-world applications where new  instances from unseen problem classes may appear after the model has been trained. This evaluation has strong implications regarding the transferability and robustness of the predictor model across diverse problem domains. To test this, we trained one individual model (one for each of the 12 problem classes) on all available problem classes except for one, which was withheld for the final validation. The results of this "leave-one-class-out" analysis are presented in Fig. \ref{fig:out-of-distribution}.

Our findings indicate that certain problem classes are highly generalizable. For example, the Knapsack, Number Partition, and Subset Sum problems exhibit low MAE of less than $7\%$, even though the model was not explicitly trained on them. This high performance suggests that the training dataset contains other problem classes with similar structural characteristics, allowing the model to transfer its learning effectively. In contrast, the model struggles to generalize to other classes, such as Weighted Max-Cut and Maximum Independent Set, where the MAE reaches values above $30\%$. While this highlights a current limitation of the architecture, the result is expected given the unique structural properties of these specific optimization problems. These findings emphasize the necessity of training the GNN on large, diverse datasets that include a wide variety of problem classes and instance types to ensure robust performance across the entire combinatorial optimization landscape.

\subsection{Finetuning}\label{sec:finetuning}

\begin{figure}[tbp]
  \centering
  \includegraphics[width=0.65\linewidth]{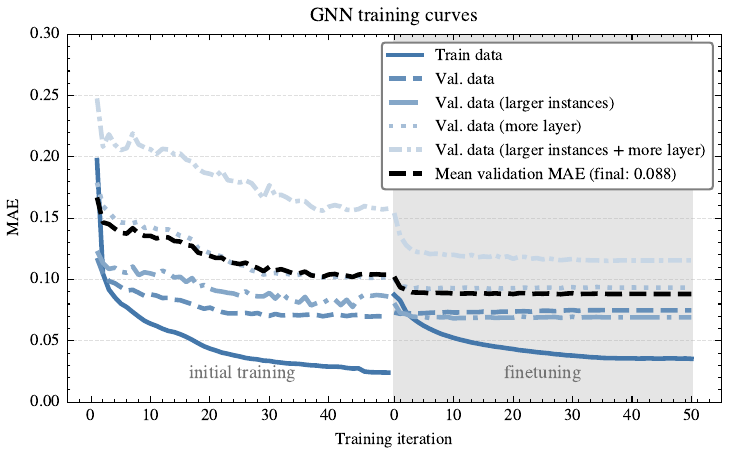}
  \caption{QAOA-Predictor's generalization over problem classes. A GNN-based model is trained on all but one of the problem classes (for all problem classes) and then the mean MAE over the validation dataset is shown for the unseen problem class.}
  \label{fig:finetuning}
  
\end{figure}

Another important feature for the predictor model to be deployed into an optimization pipeline for real world problems is the ability to finetune the model with problem instances which were not contained in the initial training data and which are not available to a high extent. This is necessary to compensate for the deteriorating performance of the  predictor model on instances of increasing size and layer count. For industrially relevant problem sizes, no simulation data is available, with only highly expensive samples from quantum hardware available for training. We therefore mimic this scenario by finetuning the GNN with a smaller dataset. The result of the finetuning can be seen in Fig. \ref{fig:finetuning}. While the original validation loss stagnates during the finetuning period, the validation loss for larger instances with more layers is decreased. However, even after this reduction, the resulting MAE for these larger and deeper instances remains substantially higher than that achieved on smaller and shallower validation instances. This observation indicates that the limited amount of fine-tuning data available is insufficient to fully compensate for their absence during the initial training phase.

\section{Conclusions and Future Work}

In this work, we introduced QAOA-Predictor, a GNN-based model designed to predict the performance of LR-QAOA across a diverse set of binary combinatorial optimization problems. Through systematic comparison with a convolutional-based model and a k-nearest neighbors baseline, we demonstrate that all models achieve high prediction accuracy. Notably, the GNN exhibited robust generalization and extrapolation capabilities, successfully predicting performance for larger problem instances and deeper circuits beyond those included during training. 

This work proposes a machine learning model capable of predicting LR-QAOA success probability, offering several practical benefits: Most importantly, it empowers users to select the minimum number of layers $p$ and the precise initialization of $\Delta_{\gamma,\beta}$ required to meet a specific performance target. By identifying the shallowest possible circuit depth, our approach ensures high-quality results while significantly conserving quantum computing resources. Furthermore, QAOA-Predictor allows users to quickly determine — without the need for quantum hardware or time-consuming simulations — whether a specific problem instance is appropriate for the LR-QAOA framework.

Future research could expand this approach to encompass a broader range of problem classes, layer depths, and parameter configurations. An especially interesting direction would be to increase problem sizes to the point where classical simulation is no longer feasible, evaluating whether the GNN's extrapolation capabilities hold in the truly large-scale regime. Additionally, training the model on data generated from quantum hardware would be a significant step toward making this method viable for industrial applications, where noise and hardware-specific constraints play a critical role. Finally, another interesting future direction would be to use QAOA-Predictor as an analytical tool to gain deeper understanding of the specific problem features that govern LR-QAOA performance, and to investigate whether tailored preprocessing techniques derived from these insights could further enhance the algorithm’s solution quality.

\paragraph{Funding Declaration}
The research is part of the Munich Quantum Valley, which
is supported by the Bavarian state government with funds from
the Hightech Agenda Bayern Plus.
%
%

\FloatBarrier

\bibliography{bib}

@article{shor1999polynomial,
  title={Polynomial-time algorithms for prime factorization and discrete logarithms on a quantum computer},
  author={Shor, Peter W},
  journal={SIAM review},
  DOI = {https://epubs.siam.org/doi/10.1137/S0097539795293172},
  volume={41},
  number={2},
  pages={303--332},
  year={1999},
  publisher={SIAM}
}

@article{egger2021warm,
  title={Warm-starting quantum optimization},
  author={Egger, Daniel J and Mare{\v{c}}ek, Jakub and Woerner, Stefan},
  journal={Quantum},
  volume={5},
  pages={479},
  year={2021},
  DOI={https://doi.org/10.22331/q-2021-06-17-479},
  publisher={Verein zur F{\"o}rderung des Open Access Publizierens in den Quantenwissenschaften}
}

@article{grimsley2019adaptive,
  title={An adaptive variational algorithm for exact molecular simulations on a quantum computer},
  author={Grimsley, Harper R and Economou, Sophia E and Barnes, Edwin and Mayhall, Nicholas J},
  journal={Nature communications},
  volume={10},
  number={1},
  pages={3007},
  year={2019},
  DOI={https://doi.org/10.1038/s41467-019-10988-2},
  publisher={Nature Publishing Group UK London}
}

@Inbook{Powell1994,
author="Powell, M. J. D.",
editor="Gomez, Susana
and Hennart, Jean-Pierre",
title="A Direct Search Optimization Method That Models the Objective and Constraint Functions by Linear Interpolation",
bookTitle="Advances in Optimization and Numerical Analysis",
year="1994",
publisher="Springer Netherlands",
address="Dordrecht",
pages="51--67",
isbn="978-94-015-8330-5",
doi="10.1007/978-94-015-8330-5_4"
}

@inproceedings{spall1987stochastic,
  title={A stochastic approximation technique for generating maximum likelihood parameter estimates},
  author={Spall, James C},
  booktitle={1987 American control conference},
  pages={1161--1167},
  year={1987},
  organization={IEEE},
   doi={10.23919/ACC.1987.4789489}}

@article{gleissner2024restricted,
  title={Restricted global optimization for QAOA},
  author={Glei{\ss}ner, Peter and Kruse, Georg and Ro{\ss}kopf, Andreas},
  journal={APL Quantum},
  DOI = {https://doi.org/10.1063/5.0189374},
  volume={1},
  number={2},
  year={2024},
  publisher={AIP Publishing}
}

@article{zaheer2017deep,
  title={Deep sets},
  author={Zaheer, Manzil and Kottur, Satwik and Ravanbakhsh, Siamak and Poczos, Barnabas and Salakhutdinov, Russ R and Smola, Alexander J},
  journal={Advances in neural information processing systems},
  DOI = {10.48550/arXiv.1703.06114},
  volume={30},
  year={2017}
}

@article{mavroeidis2018impact,
  title={The impact of quantum computing on present cryptography},
  author={Mavroeidis, Vasileios and Vishi, Kamer and Zych, Mateusz D and J{\o}sang, Audun},
  journal={arXiv preprint arXiv:1804.00200},
DOI = {10.14569/IJACSA.2018.090354},
  year={2018}
}

@article{preskill2018quantum,
  title={Quantum computing in the \uppercase{NISQ} era and beyond},
  author={Preskill, John},
  journal={Quantum},
  volume={2},
  pages={79},
  year={2018},
DOI = {https://doi.org/10.22331/q-2018-08-06-79},
  publisher={Verein zur F{\"o}rderung des Open Access Publizierens in den Quantenwissenschaften}
}

@article{farhi2014quantum,
  title={A quantum approximate optimization algorithm},
  author={Farhi, Edward and Goldstone, Jeffrey and Gutmann, Sam},
  journal={arXiv preprint arXiv:1411.4028},
DOI = {
https://doi.org/10.48550/arXiv.1411.4028},
  year={2014}
}

@article{montanez2025towards,
  title={Toward a linear-ramp QAOA protocol: evidence of a scaling advantage in solving some combinatorial optimization problems},
  author={Montanez-Barrera, JA and Michielsen, Kristel},
  journal={npj Quantum Information},
  volume={11},
  number={1},
  pages={131},
  year={2025},
  publisher={Nature Publishing Group UK London},
  DOI={https://doi.org/10.1038/s41534-025-01082-1}
}

@article{glover2018tutorial,
  title={A tutorial on formulating and using QUBO models},
  author={Glover, Fred and Kochenberger, Gary and Du, Yu},
  journal={arXiv preprint arXiv:1811.11538},
DOI = {https://doi.org/10.48550/arXiv.1811.11538},
  year={2018}
}

@article{crooks2019gradients,
  title={Gradients of parameterized quantum gates using the parameter-shift rule and gate decomposition},
  author={Crooks, Gavin E},
  journal={arXiv preprint arXiv:1905.13311},
  year={2019},
DOI = {https://doi.org/10.48550/arXiv.1905.13311}
}

@article{mcclean2018barren,
  title={Barren plateaus in quantum neural network training landscapes},
  author={McClean, Jarrod R and Boixo, Sergio and Smelyanskiy, Vadim N and Babbush, Ryan and Neven, Hartmut},
  journal={Nature communications},
  volume={9},
  number={1},
  pages={4812},
  year={2018},
  publisher={Nature Publishing Group UK London},
DOI = {https://doi.org/10.1038/s41467-018-07090-4}
}

@article{larocca2025barren,
  title={Barren plateaus in variational quantum computing},
  author={Larocca, Martin and Thanasilp, Supanut and Wang, Samson and Sharma, Kunal and Biamonte, Jacob and Coles, Patrick J and Cincio, Lukasz and McClean, Jarrod R and Holmes, Zo{\"e} and Cerezo, Marco},
  journal={Nature Reviews Physics},
  pages={1--16},
  year={2025},
  publisher={Nature Publishing Group UK London},
DOI = {https://doi.org/10.1038/s42254-025-00813-9}
}

@article{xue2021effects,
  title={Effects of quantum noise on quantum approximate optimization algorithm},
  author={Xue, Cheng and Chen, Zhao-Yun and Wu, Yu-Chun and Guo, Guo-Ping},
  journal={Chinese Physics Letters},
  volume={38},
  number={3},
  pages={030302},
  year={2021},
  publisher={IOP Publishing},
DOI = {10.1088/0256-307X/38/3/030302}
}

@article{vcepaite2025quantum,
  title={Quantum-Enhanced Optimization by Warm Starts},
  author={{\v{C}}epait{\.e}, Ieva and Vaishnav, Niam and Zhou, Leo and Montanaro, Ashley},
  journal={arXiv preprint arXiv:2508.16309},
  year={2025},
DOI = {https://doi.org/10.48550/arXiv.2508.16309}
}

@article{montanez2025transfer,
  title={Transfer learning of optimal QAOA parameters in combinatorial optimization},
  author={Montanez-Barrera, JA and Willsch, Dennis and Michielsen, Kristel},
  journal={Quantum Information Processing},
  volume={24},
  number={5},
  pages={129},
  year={2025},
  publisher={Springer},
DOI = {https://doi.org/10.1007/s11128-025-04743-4}
}

@inproceedings{pelofske2023high,
  title={High-Round QAOA for MAX $ k $-SAT on Trapped Ion NISQ Devices},
  author={Pelofske, Elijah and B{\"a}rtschi, Andreas and Golden, John and Eidenbenz, Stephan},
  booktitle={2023 IEEE International Conference on Quantum Computing and Engineering (QCE)},
  volume={1},
  pages={506--517},
  year={2023},
  organization={IEEE},
DOI = {10.1109/QCE57702.2023.00064}
}

@article{goemans1995improved,
  title={Improved approximation algorithms for maximum cut and satisfiability problems using semidefinite programming},
  author={Goemans, Michel X and Williamson, David P},
  journal={Journal of the ACM (JACM)},
  volume={42},
  number={6},
  pages={1115--1145},
  year={1995},
  publisher={ACM New York, NY, USA},
DOI = {https://doi.org/10.1145/227683.227684}
}

@article{wang2018quantum,
  title={Quantum approximate optimization algorithm for MaxCut: A fermionic view},
  author={Wang, Zhihui and Hadfield, Stuart and Jiang, Zhang and Rieffel, Eleanor G},
  journal={Physical Review A},
  volume={97},
  number={2},
  pages={022304},
  year={2018},
  publisher={APS},
DOI = {http://dx.doi.org/10.1103/PhysRevA.97.022304}
}

@article{lyngfelt2025symmetry,
  title={Symmetry-informed transferability of optimal parameters in the quantum approximate optimization algorithm},
  author={Lyngfelt, Isak and Garc{\'\i}a-{\'A}lvarez, Laura},
  journal={Physical Review A},
  volume={111},
  number={2},
  pages={022418},
  year={2025},
  publisher={APS},
DOI = {https://doi.org/10.1103/PhysRevA.111.022418}
}

@article{bergholm2018pennylane,
  title={Pennylane: Automatic differentiation of hybrid quantum-classical computations},
  author={Bergholm, Ville and Izaac, Josh and Schuld, Maria and Gogolin, Christian and Ahmed, Shahnawaz and Ajith, Vishnu and Alam, M Sohaib and Alonso-Linaje, Guillermo and AkashNarayanan, B and Asadi, Ali and others},
  journal={arXiv preprint arXiv:1811.04968},
  year={2018},
DOI = {https://doi.org/10.48550/arXiv.1811.04968}
}

@article{apolloni1989quantum,
  title={Quantum stochastic optimization},
  author={Apolloni, Bruno and Carvalho, C and De Falco, Diego},
  journal={Stochastic Processes and their Applications},
  volume={33},
  number={2},
  pages={233--244},
  year={1989},
  publisher={Elsevier},
DOI = {https://doi.org/10.1016/0304-4149(89)90040-9}
}

@article{scarselli2008graph,
  title={The graph neural network model},
  author={Scarselli, Franco and Gori, Marco and Tsoi, Ah Chung and Hagenbuchner, Markus and Monfardini, Gabriele},
  journal={IEEE transactions on neural networks},
  volume={20},
  number={1},
  pages={61--80},
  year={2008},
  publisher={IEEE},
DOI = {10.1109/TNN.2008.2005605}
}

@inproceedings{grover1996fast,
  title={A fast quantum mechanical algorithm for database search},
  author={Grover, Lov K},
  booktitle={Proceedings of the twenty-eighth annual ACM symposium on Theory of computing},
  pages={212--219},
  year={1996},
DOI = {https://doi.org/10.1145/237814.237866}
}

@article{Harrow_2009,
   title={Quantum Algorithm for Linear Systems of Equations},
   volume={103},
   ISSN={1079-7114},
   url={http://dx.doi.org/10.1103/PhysRevLett.103.150502},
   DOI={10.1103/physrevlett.103.150502},
   number={15},
   journal={Physical Review Letters},
   publisher={American Physical Society (APS)},
   author={Harrow, Aram W. and Hassidim, Avinatan and Lloyd, Seth},
   year={2009},
   month=oct }

@article{tilly2022variational,
  title={The variational quantum eigensolver: a review of methods and best practices},
  author={Tilly, Jules and Chen, Hongxiang and Cao, Shuxiang and Picozzi, Dario and Setia, Kanav and Li, Ying and Grant, Edward and Wossnig, Leonard and Rungger, Ivan and Booth, George H and others},
  journal={Physics Reports},
  volume={986},
  pages={1--128},
  year={2022},
  publisher={Elsevier},
DOI = {https://doi.org/10.1016/j.physrep.2022.08.003}
}

@article{montanez2024unbalanced,
  title={Unbalanced penalization: A new approach to encode inequality constraints of combinatorial problems for quantum optimization algorithms},
  author={Monta{\~n}ez-Barrera, Jhon Alejandro and Willsch, Dennis and Maldonado-Romo, Alberto and Michielsen, Kristel},
  journal={Quantum Science and Technology},
  volume={9},
  number={2},
  pages={025022},
  year={2024},
  publisher={IOP Publishing},
DOI = {10.1088/2058-9565/ad35e4}
}

@article{kandala2017hardware,
  title={Hardware-efficient variational quantum eigensolver for small molecules and quantum magnets},
  author={Kandala, Abhinav and Mezzacapo, Antonio and Temme, Kristan and Takita, Maika and Brink, Markus and Chow, Jerry M and Gambetta, Jay M},
  journal={nature},
  volume={549},
  number={7671},
  pages={242--246},
  year={2017},
  publisher={Nature Publishing Group},
DOI = {https://doi.org/10.1038/nature23879}
}

@article{yarkoni2022quantum,
  title={Quantum annealing for industry applications: Introduction and review},
  author={Yarkoni, Sheir and Raponi, Elena and B{\"a}ck, Thomas and Schmitt, Sebastian},
  journal={Reports on Progress in Physics},
  volume={85},
  number={10},
  pages={104001},
  year={2022},
  publisher={IOP Publishing},
DOI = {10.1088/1361-6633/ac8c54}
}

@article{cerezo2021variational,
  title={Variational quantum algorithms},
  author={Cerezo, Marco and Arrasmith, Andrew and Babbush, Ryan and Benjamin, Simon C and Endo, Suguru and Fujii, Keisuke and McClean, Jarrod R and Mitarai, Kosuke and Yuan, Xiao and Cincio, Lukasz and others},
  journal={Nature Reviews Physics},
  volume={3},
  number={9},
  pages={625--644},
  year={2021},
  publisher={Nature Publishing Group UK London},
DOI = {https://doi.org/10.1038/s42254-021-00348-9

}
}

\begin{appendices}

\section{}\label{secA1}

\subsection{Dataset}\label{secA:dataset}

\begin{figure}[ht]
\centering

\includegraphics[width=0.305\linewidth]{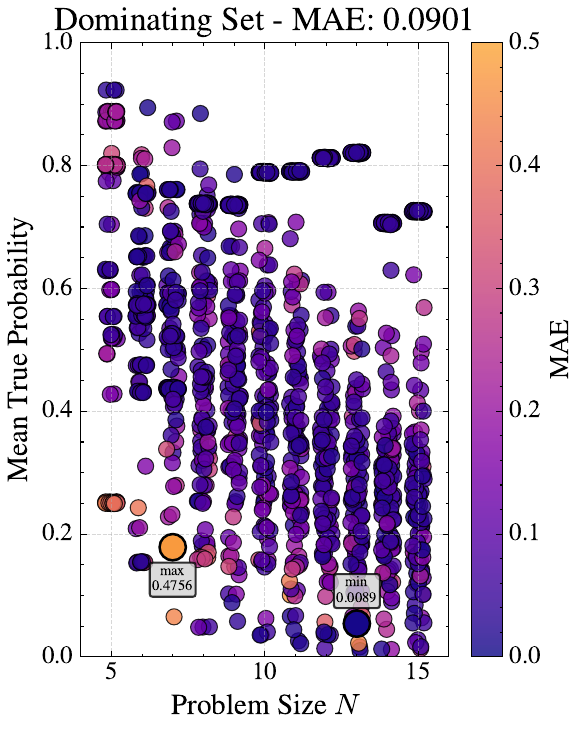}\hfill
\includegraphics[width=0.305\linewidth]{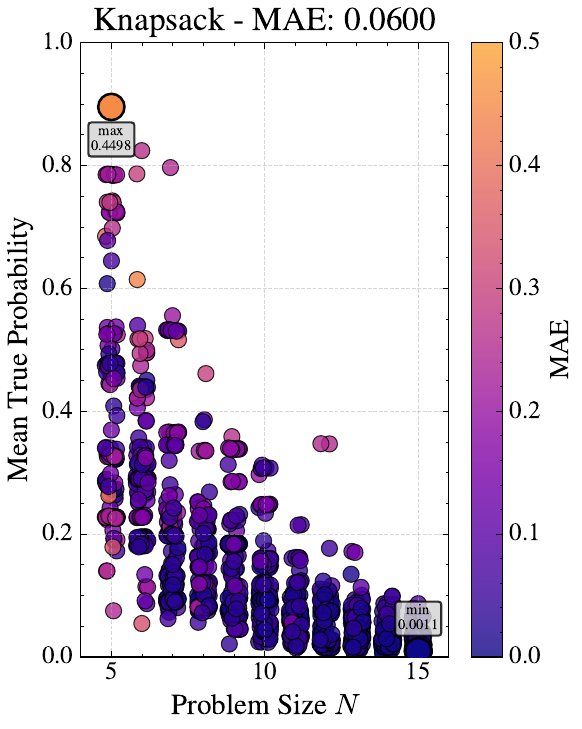}\hfill
\includegraphics[width=0.305\linewidth]{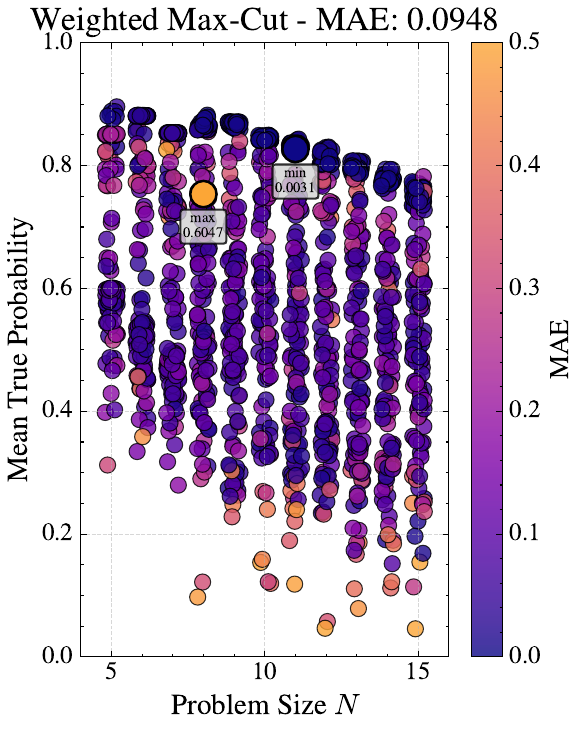}

\vspace{1pt}

\includegraphics[width=0.305\linewidth]{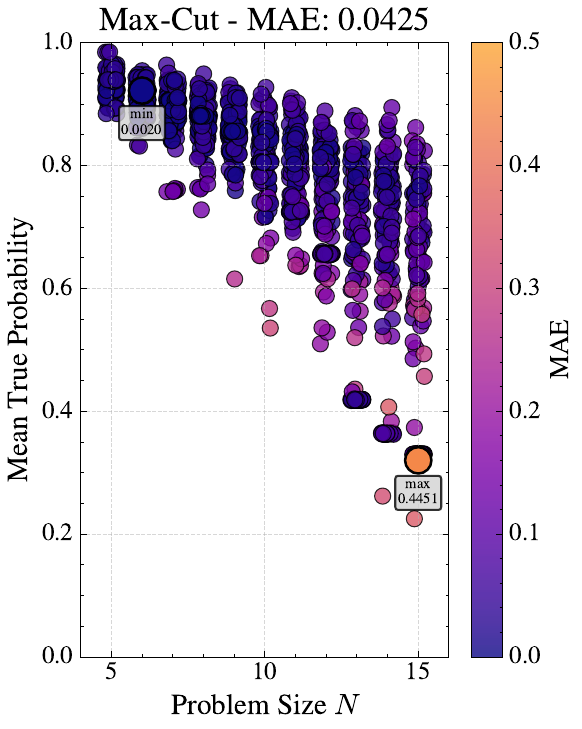}\hfill
\includegraphics[width=0.305\linewidth]{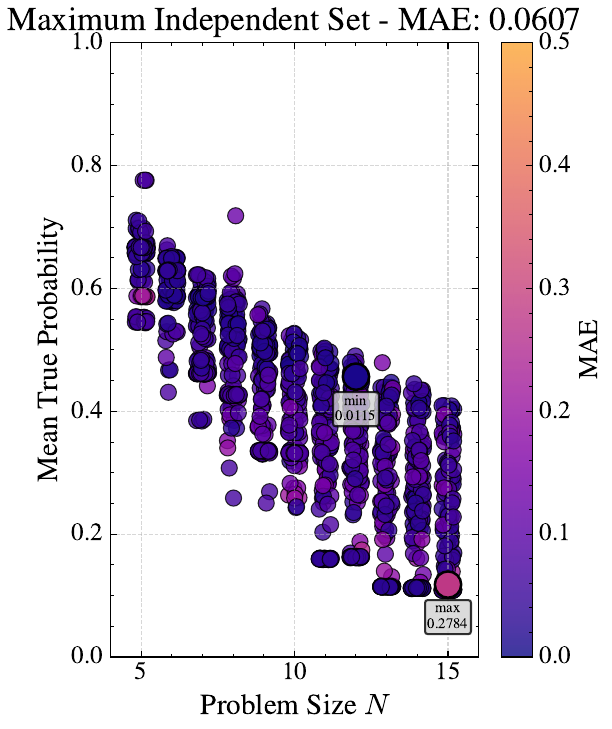}\hfill
\includegraphics[width=0.305\linewidth]{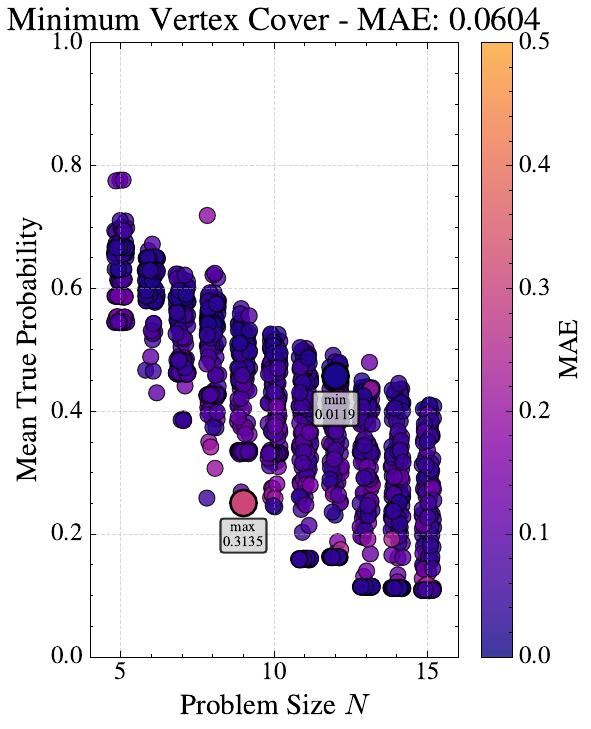}
\caption{Example instances of the first half of the 12 problem classes.}
\label{fig:scatter_plots1}
\end{figure}

\begin{figure}[!t]
\centering
\includegraphics[width=0.305\linewidth]{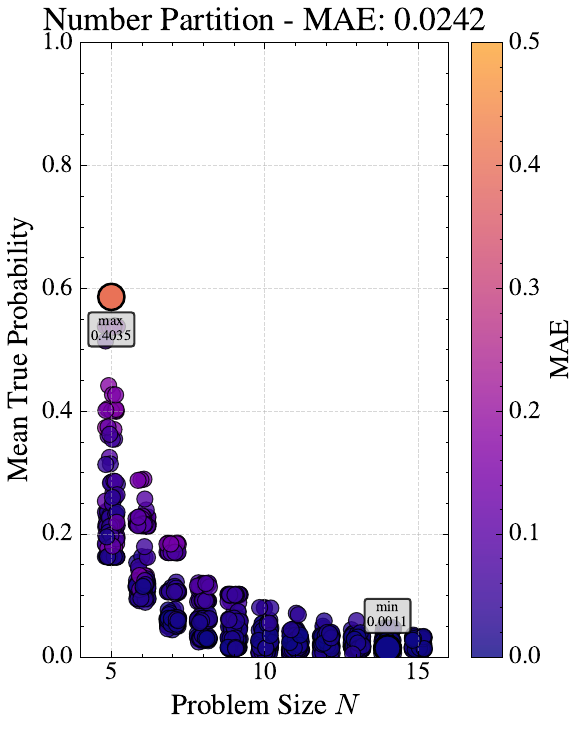}\hfill
\includegraphics[width=0.305\linewidth]{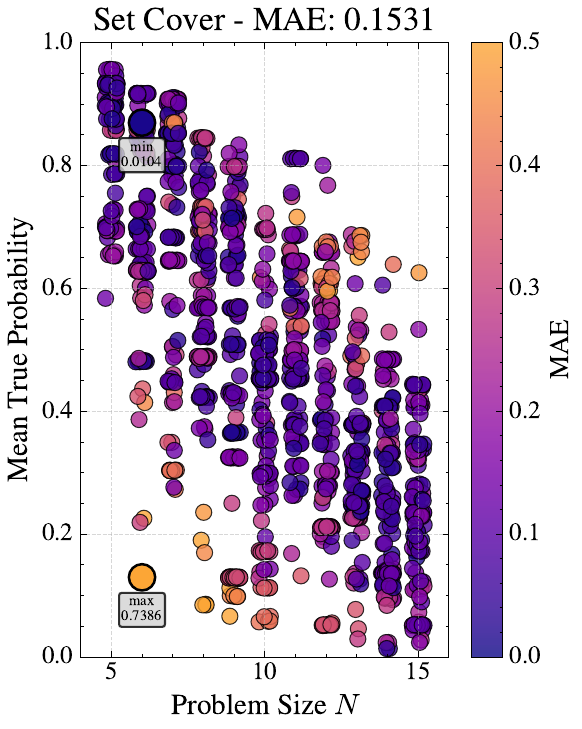}\hfill
\includegraphics[width=0.305\linewidth]{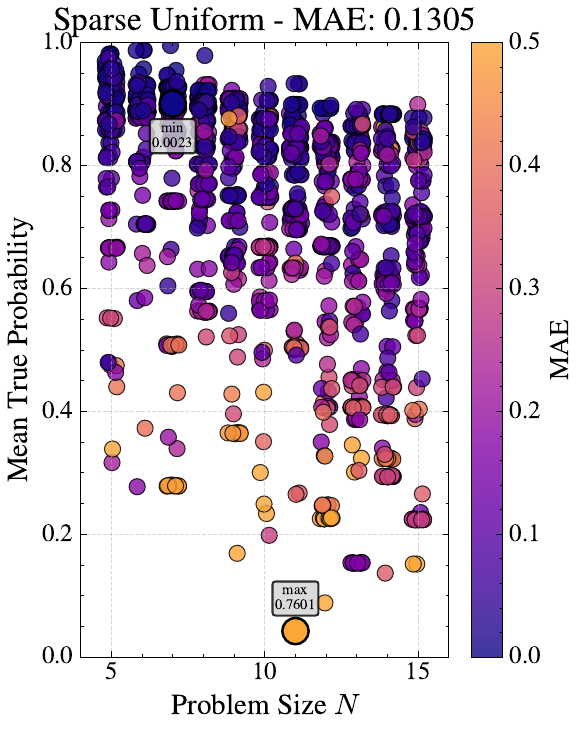}

\vspace{1pt}

\includegraphics[width=0.305\linewidth]{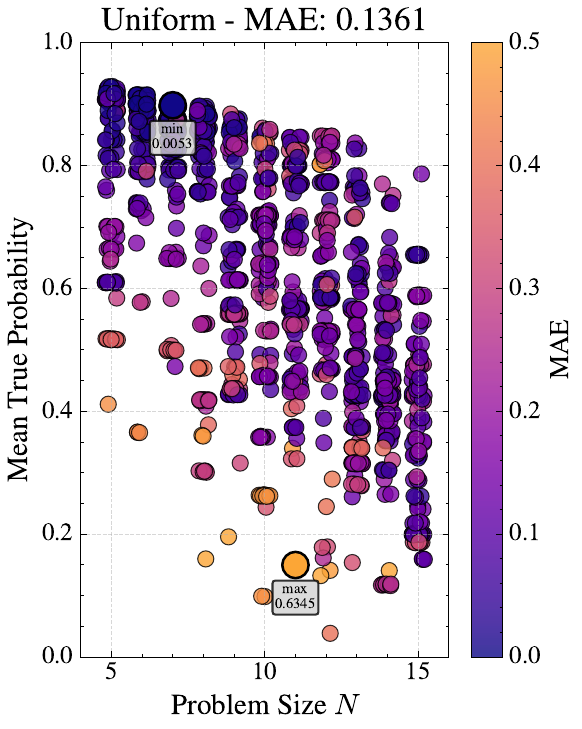}\hfill
\includegraphics[width=0.305\linewidth]{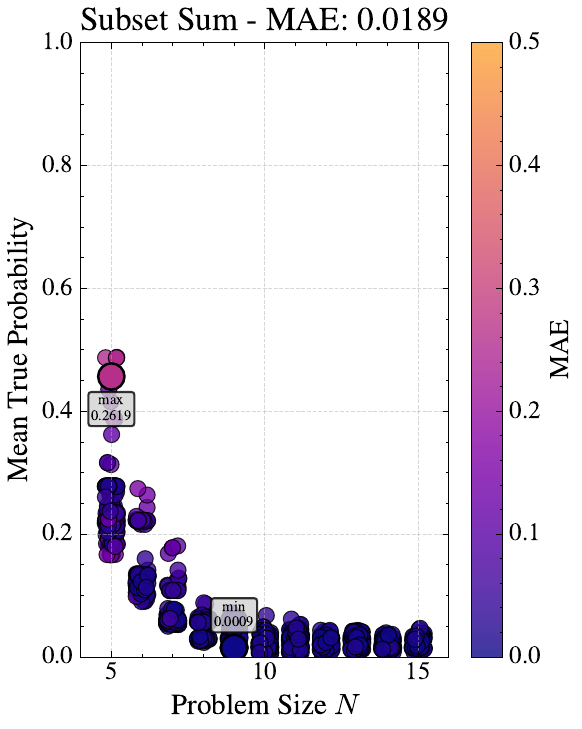}\hfill
\includegraphics[width=0.305\linewidth]{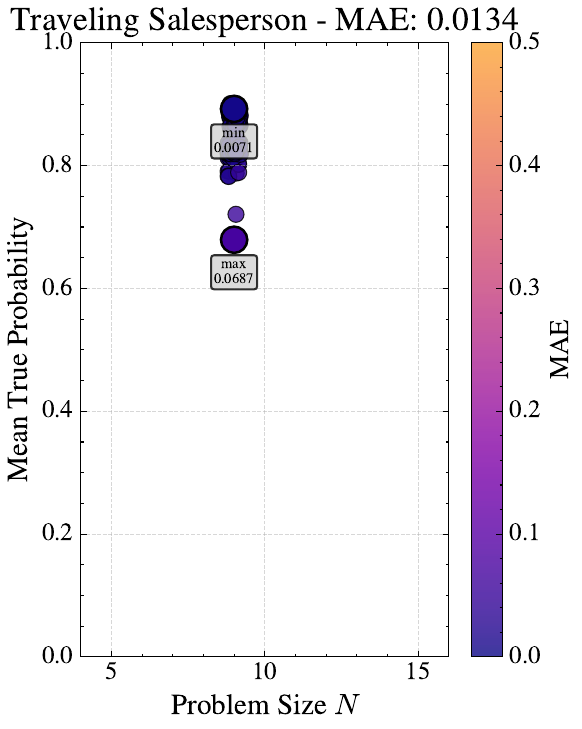}

\caption{Example instances of the second half of the 12 problem classes.}
\label{fig:scatter_plots2}
\end{figure}

\begin{figure}[!t]
\centering

\includegraphics[width=0.305\linewidth]{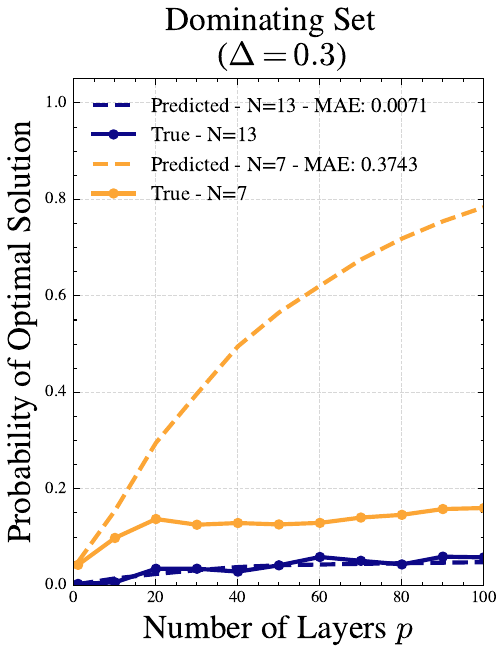}\hfill
\includegraphics[width=0.305\linewidth]{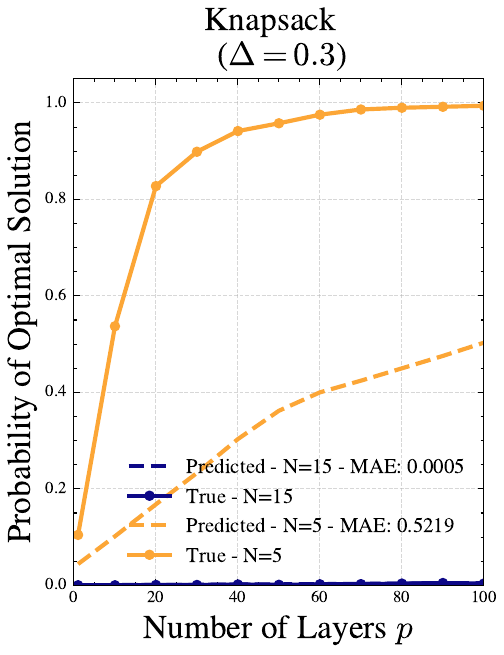}\hfill
\includegraphics[width=0.305\linewidth]{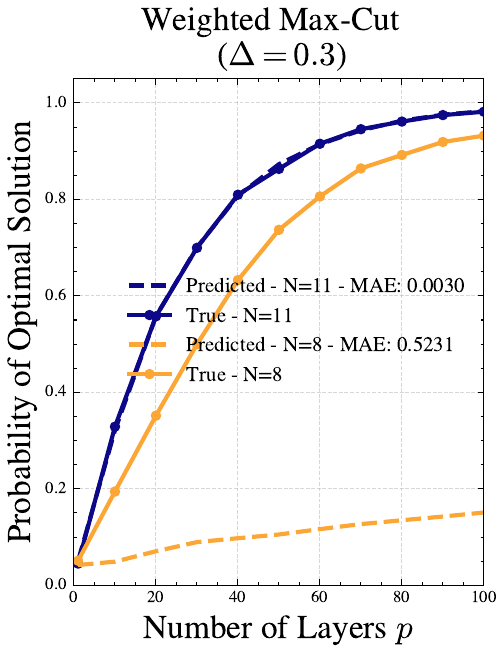}

\vspace{1pt}

\includegraphics[width=0.305\linewidth]{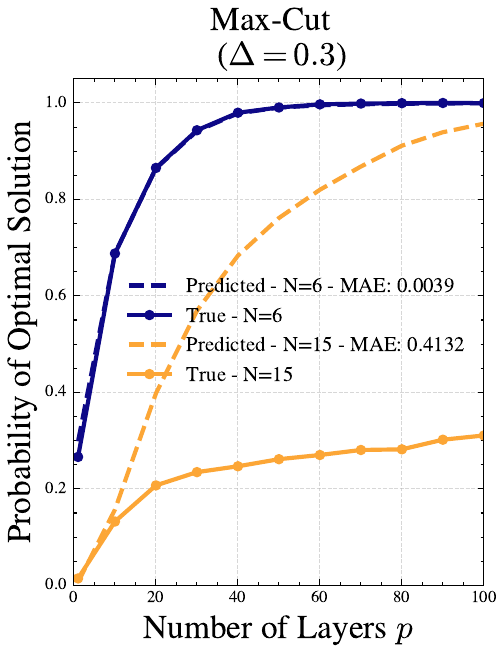}\hfill
\includegraphics[width=0.305\linewidth]{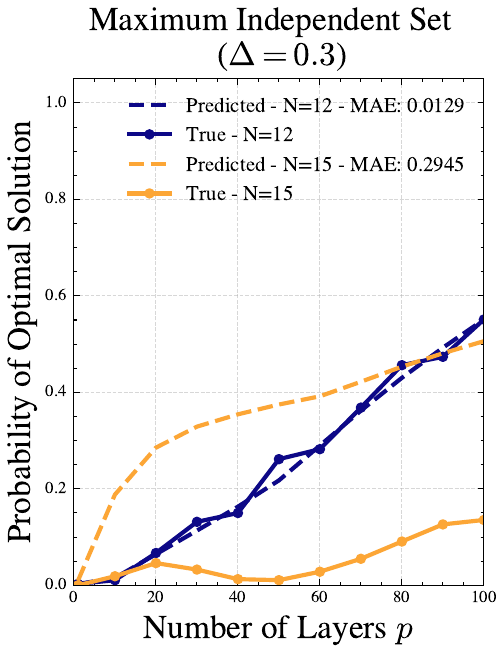}\hfill
\includegraphics[width=0.305\linewidth]{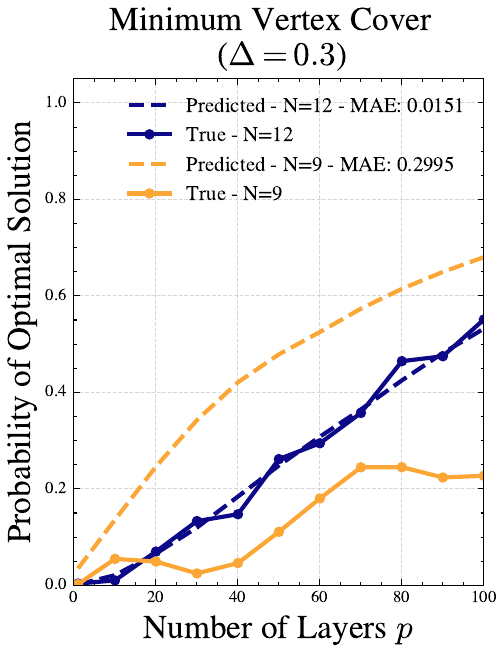}

\caption{Example instances of the first half of the 12 problem classes.}
\label{fig:line_plots1}
\end{figure}

\begin{figure}[!t]
\centering
\includegraphics[width=0.305\linewidth]{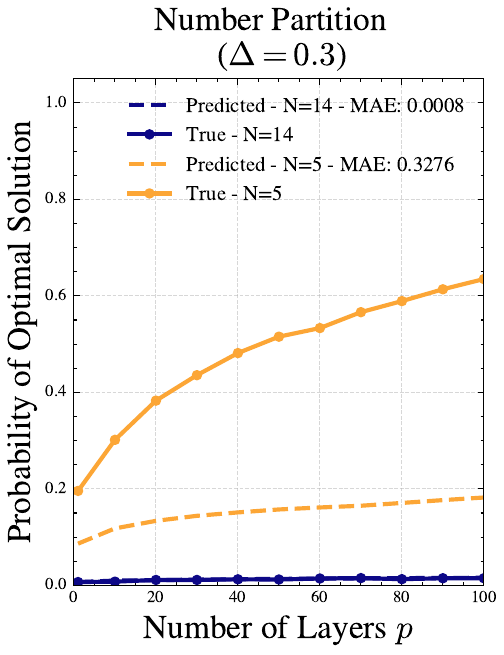}\hfill
\includegraphics[width=0.305\linewidth]{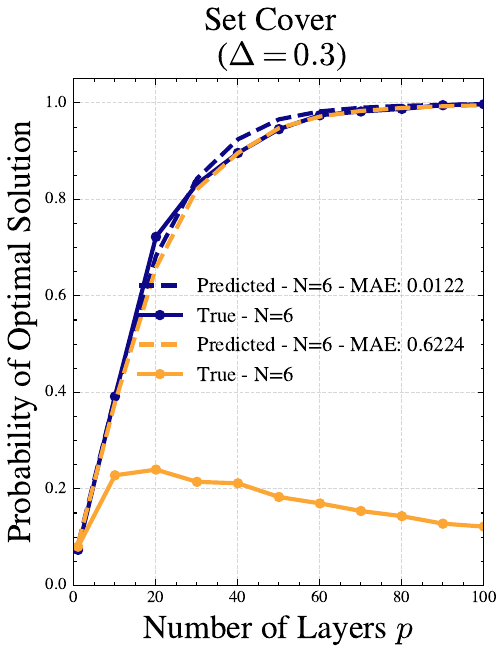}\hfill
\includegraphics[width=0.305\linewidth]{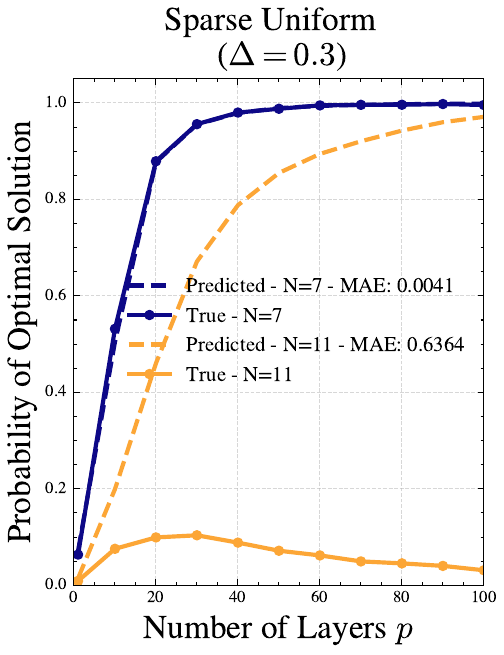}

\vspace{1pt}

\includegraphics[width=0.305\linewidth]{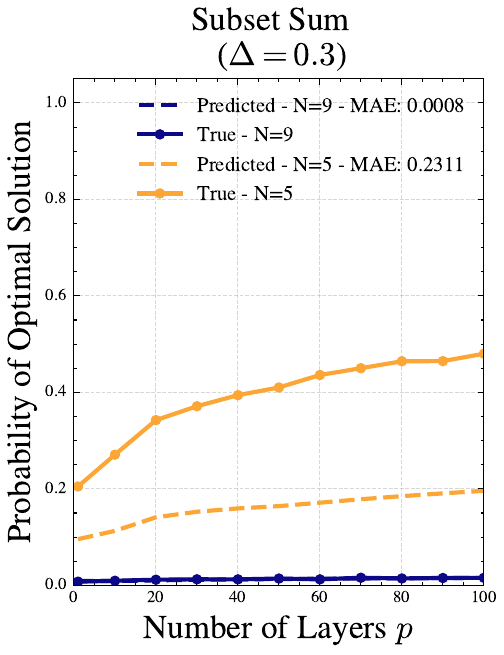}\hfill
\includegraphics[width=0.305\linewidth]{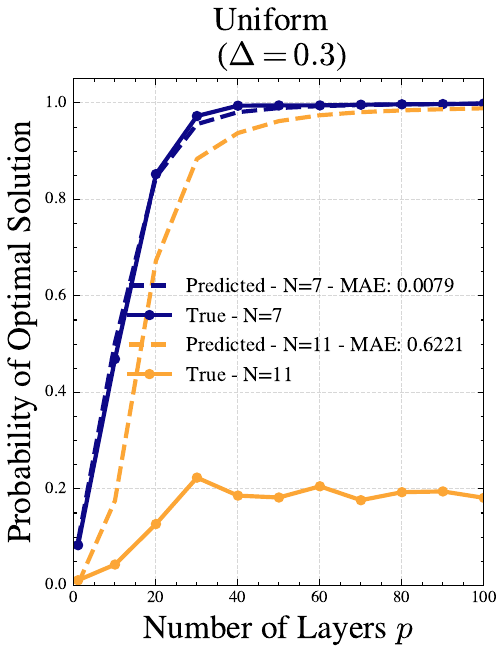}\hfill
\includegraphics[width=0.305\linewidth]{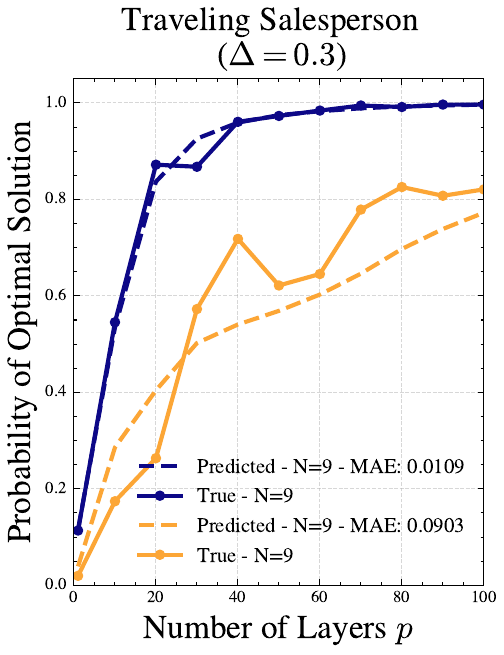}

\caption{Example instances of the second half of the 12 problem classes.}

\label{fig:line_plots2}
\end{figure}

\begin{table}[htbp]
\centering
\caption{Complete composition of the synthetic dataset. We distinguish between graph based and non-graph based problem classes. For the graph-based set, $I$ instances are generated for each combination of problem class, graph type, graph density and instance size $n$. For the non-graph based set, $I\times C$ instances are generated for each combination of problem class and instance size $n$. Here, $I=50$ and $I=10$ for the training and validation datasets, respectively. Moreover, $C$ is a constant equal to the number of graph types times the number of graph densities used to create the graph based instances to assert that we have an equal number of graph and non-graph based instances on each dataset. The Traveling Salesperson is only included for $N=9$ (so $750$ and $150$ instances for training and validation dataset respectively). Finally, these instances do not correspond to the final data points, since the models (like the GNN) receive not only a QUBO matrix/graph corresponding to a specific instance but also specific values of $\Delta_{\gamma,\beta}$ and $p$. If not explicitly stated otherwise, the same values are used for training and validation dataset. Differing values are indicated by $training/validation$ The finetuning dataset is generated as the validation dataset, except that $I=1$ (so $10\%$ of the size).}
\label{tab:dataset}
\begin{tabular}{ll}
\toprule
\textbf{Parameter}              & \textbf{Values}                               \\
\midrule
Instance size $n$               & $\{5,\dots,15\}$ (11 values)  / validation $\{5, \dots , 20\}$ (16 values)              \\
Ramp parameter $\Delta_{\gamma,\beta}$   & $\{0.3, 0.5\}$ (2 values)                     \\
Number of layers $p$        & $\{1,10,20,\dots,100\}$ (11 values) / validation $\{1,\dots,100,150,250,500\}$ (14 values)             \\
\midrule
Graph problem classes       & Max-Cut, Weighted Max-Cut, MVC, MIS,            \\
                                & Dominating Set, Traveling Salesperson (6 values)   \\
Graph Types                     & \{Erdos Renyi, Barabasi Albert, Complete Bipartite\} (3 values) \\
Graph densities                 & $\{0.1,0.25,0.5,0.75,0.9\}$ (5 values)  \\
$\rightarrow$ Graph instances   & $50/10\times 11/16 \times 5 \times 3 \times 5 + (50/10\times3\times5) = 42{,}000$ / validation $12{,}150$\\
$\rightarrow$ DataPoints   & $42{,}000/12{,}150 \times 2\times11/14=924{,}000$ / validation $340{,}200$\\
\midrule
Non-graph problem classes   & Knapsack, Uniform, Sparse Uniform,            \\
                                & Number Partition, Subset Sum, Set Cover  (6 values)     \\
$\rightarrow$ Non-graph instances & $50/10 \times 11/16 \times 15 \times 6 = 49{,}500$ / validation $14{,}400$\\
$\rightarrow$ DataPoints   & $49{,}500/14{,}400 \times 2\times11/14=1{,}089{,}000$ / validation $403{,}200$\\
\midrule
\textbf{Total Training}                  & \textbf{91,500 instances and 2,013,000 datapoints}                    \\
\textbf{Total Validation}                  & \textbf{26,550 instances and 743,400 datapoints}                    \\
\textbf{Total Finetuning}                  & \textbf{2,655 instances and 74,340 datapoints ($\leftarrow$ 10\% of validation dataset)}                    \\
\bottomrule
\end{tabular}
\end{table}

\FloatBarrier

\subsection{Hyperparameter Tuning}\label{secA:hyperparameters}

\begin{table}[htbp]
\centering
\caption{Hyperparameter search space and final chosen configuration for the GNN. 
         The search was performed using grid-search over all possible combinations of parameters. To select the best hyperparameters, the mean of all validation loss terms defined in Sec.\ref{sec:tuning} was used.
         }
\label{tab:hyperparameters_GNN}
\begin{tabular}{lcc}
\toprule
\textbf{Hyperparameter}          & \textbf{Search Space}                              & \textbf{Chosen}         \\
\midrule
Hidden dimension $d$             & \{64, 128, 256, 512\}                              & 128        \\
Number of GNN layers $L$         & \{2,4,8,12\}                              & 4          \\
Readout / pooling                & Mean, Max, MeanMax, Attention                            & MeanMax        \\
Dropout rate                     & \{0.0, 0.05, 0.1, 0.25\}                        & 0.05        \\
Batch size                       & \{4096,8192,16384\}                               & 16284        \\
Learning rate (Adam)             & \{1e-5, 1e-4, 1e-3, 5e-3, 1e-2\}                    & 1e-3        \\
Weight decay                     & \{0, 1e-5, 5e-5\}                                  & 0        \\
Loss function                    & MAE, MSE                       & MAE        \\

\bottomrule
\end{tabular}
\end{table}

\begin{table}[htbp]
\centering
\caption{Hyperparameter search space and final chosen configuration for the CNN. 
         The search was performed using grid-search over all possible combinations of parameters. To select the best hyperparameters, the mean of all validation loss terms defined in Sec.\ref{sec:tuning} was used.
         }
\label{tab:hyperparameters_CNN}
\begin{tabular}{lcc}
\toprule
\textbf{Hyperparameter}          & \textbf{Search Space}                              & \textbf{Chosen}         \\
\midrule
Hidden dimension $d$             & \{64, 128, 256\}                              & 256        \\
Number of CNN layers $L$         & \{2,4\}                              & 2          \\
Dropout rate                     & \{0.0, 0.05, 0.1, 0.25\}                        & 0.05        \\
Batch size                       & \{256,512,1024,2048\}                               & 256        \\
Learning rate (Adam)             & \{1e-5, 5e-5, 1e-4, 5e-4, 1e-3, 5e-3\}                    & 5e-3        \\
Weight decay                     & \{0, 1e-5, 5e-5\}                                  & 0        \\
Loss function                    & MAE, MSE                       & MAE        \\

\bottomrule
\end{tabular}
\end{table}

\begin{table}[htbp]
\centering
\caption{Key hyperparameters and setup details of the KNN baseline.}
\label{tab:knn-params}
\begin{tabular}{ll}
\toprule
\textbf{Parameter}                  & \textbf{Value / Description} \\
\midrule
Regressor type                      & k-Nearest Neighbors (KNeighborsRegressor) \\
Number of neighbors (\(k\))         & 10 \\
Weighting scheme                    & distance (closer neighbors have higher weight) \\
Feature vector (11 dimensions)      & $n$, density, sparsity, diag\_mean, diag\_std, \\
                                    & off\_diag\_mean, off\_diag\_std, num\_edges, \\
                                    & avg\_degree, $\Delta_{\gamma, \beta}$, \(p\) \\

\bottomrule
\end{tabular}
\end{table}




\end{appendices}



\end{document}